\documentclass[%
 reprint,
superscriptaddress,https://www.overleaf.com/project/5dc3febdc6430c00018feb6d
linenumbers,
nofootinbib,
longbibliography,
amsmath,amssymb,
aps,
]{revtex4-1}

\usepackage{graphicx}
\usepackage{dcolumn}
\usepackage{bm}

\usepackage{color,colortbl}
\usepackage{units}
\usepackage{tikz}
\usetikzlibrary{arrows,shapes}

\usepackage{algorithm}
\usepackage[noend]{algpseudocode}
\usepackage{romannum}

\usepackage{diagbox} 

\newcolumntype{C}[1]{>{\centering\let\newline\\\arraybackslash\hspace{0pt}}m{#1}}

\usepackage{subfigure}

\renewcommand{\arraystretch}{1.25}

\definecolor{mygreen}{rgb}{0.1, 0.6, 0.0}

\definecolor{pap}{rgb}{0.54, 0.17, 0.89}

\newcommand{\xv}{{\bf x}}
\usepackage{blindtext}
\usepackage[inline]{enumitem}

\usepackage{hyperref}
\usepackage{float}

\DeclareMathAlphabet\mathbfcal{OMS}{cmsy}{b}{n}

\usepackage{soul}
\usepackage{xcolor,cancel}

\begin{document}
	
	\preprint{APS/123-QED}
	
	\title{Multi-objective design optimization of a Quadrupole Resonator under uncertainties}
	
	\author{Piotr Putek}
	\email{piotr.putek@uni-rostock.de}
	\affiliation{Institute of General Electrical Engineering, University of Rostock, 18059 Rostock, Germany}
	
	\author{Shahnam Gorgi Zadeh}
	\affiliation{Institute of General Electrical Engineering, University of Rostock, 18059 Rostock, Germany}
	
	\author{Marc Wenskat}
	\altaffiliation[Also at ]{Deutsches Elektronen-Synchrotron, 22607 Hamburg, Germany}
	\affiliation{Institute of Experimental Physics, Universit\"at Hamburg, 22761 Hamburg, Germany}
	
	\author{Ursula van Rienen}
	\altaffiliation[Also at ]{Department Life, Light \& Matter, University of Rostock, 18051 Rostock, Germany}
	\affiliation{Institute of General Electrical Engineering, University of Rostock, 18059 Rostock, Germany}
	
	
	\begin{abstract}
		{\color{black}
		For a precise determination of the radio frequency (RF) properties of superconducting materials, a calorimetric measurement is carried out with the aid of a so-called Quadrupole Resonator (QPR). This procedure is affected by certain systematic measurement errors with various sources of uncertainties. 
		In this paper, to reduce the impact of geometrical uncertainties on the measurement bias, the modified steepest descent method is used for the multi-objective shape optimization of a QPR {in terms of an expectation measure}. Thereby, variations of geometrical parameters are modeled by the Polynomial Chaos (PC) expansion technique. Then, the resulting Maxwell's eigenvalue problem with random input data is solved using the PC-based stochastic collocation method (PC-SCM). Furthermore, to assess the contribution of the particular geometrical parameters, the variance-based sensitivity analysis is proposed. This allows for modifying the steepest descent algorithm, which results in reducing the computational load needed to find optimal solutions. Finally,  optimization results in the form of an efficient approximation of the Pareto front for a three dimensional (3D) model of the QPR are shown.
		}
	\end{abstract}
	
	\pacs{Valid PACS appear here}
	\maketitle
	
	
	\section{Introduction}
	\label{sec:introduction}
	
	In modern particle accelerators, superconducting radio frequency (SRF) cavities are widely used to provide high accelerating gradients to a beam of particles {while ensuring moderate power losses}. In this respect, the physical features of materials used for building such cavities are of key importance as power consumption and maximum accelerating gradient are predominantly specified by the material properties, the surface resistance and the critical RF-field. Furthermore, since lower surface resistance implies reduced power consumption, there is a high demand on conducting experiments {to precisely determine the superconducting properties of such materials.}
	Thus, to explore the fundamental properties of superconducting materials that are used in modern particle accelerators, high precision surface-resistance measurements in a dedicated testing equipment are required. They are also interesting from a theoretical perspective.
    
    {The surface resistance of superconducting materials is in the range of tens of nano-ohms at very low temperature. The QPR is a special dedicated device used for the measurement of the surface resistance of superconducting samples, and is composed of a pillbox-like cavity containing four-vertically placed hollow rods~\cite{IEEEhowto:Kleindienst,IEEEhowto:Mahner,Tobias_Junginger2012}. By exciting a quadrupole-like magnetic field on the superconducting sample and using calorimetric methods, the surface resistance of the sample is investigated. The measurement data and the expected loss on the sample, which is obtained from the RF simulation of {the} QPR, are used to determine the surface resistance of the sample.} The measurement procedure is affected by various sources of uncertainties. In general, they are related to the resolution of electronic equipment, geometrical deviations of a cavity design, {and the accuracy of numerical simulations}. In addition, the surface treatment methods including ultrasonic bath, buffered chemical polishing and high pressure rinsing result in a certain level of surface roughness~\cite{IEEEhowto:Keckert}. As a result, they all have a direct impact on the accuracy of the surface-resistance measurement and therefore should be reflected in the modeling procedure in order to provide reliable and robust simulation results. 
	
	In recent years, {a considerable number of studies have been carried out} on uncertainty quantification (UQ) within the context of accelerator physics, see, e.g.,~\cite{Akcelik2007, putek:schmidt2014, corno2015isogeometric, putek2019, GEORG2019228,corno2019}. In general, methods that can be used to find the estimation of the statistical moments can be divided into two main groups. The first one corresponds to sampling methods such as the Monte Carlo-based approach~\cite{Moro95,niederreiter1978}, while the stochastic collocation and spectral Galerkin method belong to non-sampling-based techniques~\cite{Xiu02,Babu10}. Other approaches exist and are known such as the perturbation method~\cite{Harbrecht201091,Roemer2014}. 
	
    In many applications, a shape optimization problem with uncertainties is {usually} formulated in terms of objectives which contradict each other. In such a case, the Pareto concept appears, which can be understood as a set of optimal compromises between the conflicting objectives~\cite{IEEEhowto:miettinen99}.
	The major research on the multi-objective (MO) shape optimization, which are based on deterministic assumptions {(randomness of input parameters {is} not considered)}, might yield impractical or sub-optimal solutions due to the real-engineering conditions, {resulting} in various sources of uncertainties~\cite{YAO2011450}. 
	To deal with this problem, e.g., the concept of the almost Pareto-optimal points has been introduced in~\cite{White1986}. Probabilistic approaches to the multi-objective (MO) optimization with uncertainties, in turn, have been derived in~\cite{10.1007/3-540-44719-9_23, 10.1007/3-540-44719-9_22, KUGELMANN20187, pute15a}. Furthermore, to compute a robust approximation of the Pareto set, evolutionary algorithms have been developed for optimization and/or inverse problems with either uncertain or noise-corrupted input data~\cite{deb:09,Basseur}. In~\cite{Marzouk2009} an efficient numerical strategy for the Bayesian solution of inverse problems based on the Polynomial Chaos (PC)-based stochastic collocation method (PC-SCM) has been used to construct a polynomial approximation of the forward solution over the support of the prior distribution. Correspondingly, the gradient-based MO optimization method with uncertainties has been successfully designed and used in~\cite{Peitz2018}. In~\cite{IEEEhowto:pute18}, in turn, the physically-justified weighted sum method has been explored to solve the robust shape optimization problem, constrained by a stochastic partial differential equation (PDE). Within {that} work, the MO shape optimization problem under uncertainties has been reformulated in terms of the variance-based sensitivity (VBS) analysis~\cite{IEEEhowto:sobol, IEEEhowto:sudret2008}.
	
    In this setting, SRF cavities require an advanced, simulation-based approach to attain such a design, which satisfies demanding performance requirements in view of uncertainties due to manufacturing imperfections. In the context of accelerator physics, work by~\cite{putek:krancjevic_zadeh2019} appears to be an initial attempt to cope with the MO shape optimization problem of an SRF accelerating cavity, when considering robustness against geometric perturbations. To account for the robustness of the cavity, the sum of local partial derivatives (of some figures of merit of the cavity with respect to the design variables) were considered as an objective function in the formulation of the MO optimization. In such a formulation, the sum of partial derivatives serves as an estimator of the variance of the objective functions if the design variables are locally perturbed. Additionally, to deal with the large complexity of the MO shape optimization, due to the used genetic algorithm, the concept of the global sensitivity has been applied.  
    
    In the framework of QPR optimization, a physically-based method, developed at the Helmholtz-Zentrum  Berlin (HZB), has been proposed in~\cite{IEEEhowto:Kleindienst}. In this work, the first-order approximation of partial derivatives {of objective functions, obtained from parameter sweep,} has been used for modifying the QPR {design}. This approach, however, allows only for finding a better design in the average sense. It means that the improvement of all considered conflicting objectives is not guaranteed (even without introducing the uncertain input parameters). Though, this modified design of a resonant cavity had not been optimal from the mathematical view point, it was quite successful. It should be noticed, however, that these approaches, which eventually are based on deterministic assumptions, need to be carefully applied, since the cavity shape significantly influences the eigenmodes and eigenvectors as well as other figures of merit. For this reason, a local measure in the form of partial derivatives may not provide reliable results for both the forward and the optimization problem~\cite{corno2019}.
	
	{The main contribution of this paper is, on the one hand, to optimize the existing HZB-QPR under geometric uncertainties in order to increase the average magnetic field on the sample which consequently leads to a better measurement resolution. On the other hand, equally important objectives are to increase the homogeneity of the magnetic field distribution on the sample and also to reduce the field within the coaxial gap, which results in decreasing the unwanted heating of {the} normal-conducting flange which helps to mitigate the measurement bias, observed for the third mode. For these reasons, the shape optimization problem is formulated in terms of the expected values of suitably chosen figures of merit. Hereby, in order to mimic the production tolerances, the concept of the UQ is involved in the modeling phase of the QPR.}
	The crucial achievement of this work lies also in incorporating the variance-based sensitivity into the MO optimization formulation. This results in further reducing the computational burden, needed for efficient approximation of the Pareto front. According to the author’s best knowledge, the MO shape optimization problem of the QPR with geometrical random input parameters has not been studied yet in the proposed framework.
	
	The structure of the paper is as follows: Section~\ref{sec:phys-model} focuses on the physical model of the QPR, the calorimetric method and corresponding systematic errors and various sources of uncertainties.   
	Section~\ref{sec:StochforwProb}, in turn, deals with the stochastic Maxwell's eigenvalue problem. The UQ via the PC-SCM including the concept of the variance-based decomposition is a topic of Section~\ref{sec:PseuSpectrAppr}. 
	In Section~\ref{sec:moso}, the constrained MO problem with uncertainties is formulated and solved. Furthermore, the efficiency of the proposed method is shown in Section~\ref{sec:NumResults}. 
	Finally, Section~\ref{sec:cons} involves concluding remarks and promising directions of ongoing research.

	\section{Physical model of a QPR} 
	\label{sec:phys-model}
	
	In this section, the structure and operating principles of a QPR are shortly discussed. Several figures of merit are revisited, which allow for assessing different geometries. The uncertainties related to the calorimetric method and the design of a QPR are briefly reviewed in order to illustrate the motivation for our work. 
    
    \subsection{Mechanical Design}

    In the particle accelerator technology, current RF resonators are most often made of niobium and operated at very low temperatures in a superconducting state to minimize surface losses. 
    For a precise determination of the RF properties of such superconducting materials, QPRs are used, exploiting a well-known calorimetric measurement method~\cite{IEEEhowto:Mahner}. 
    {There are also other design for test resonators to obtain RF properties, see~\cite{osti_1329797} for an overview. However, the QPR design has two advantages (i) it allows a direct measurement of the surface resistance with a sub-n$\Omega$ resolution (ii) the applied temperature, frequency and magnetic field values are of the typical range for accelerator operation. Hence, the advantages allow to directly relate obtained results to cavity results.}
    Its design was originally developed in 1990's at CERN~\cite{CernQPR90s} and was further adopted to \unit[433]{MHz}, \unit[866]{MHz} and \unit[1.3]{GHz}, respectively, at the HZB. {Furthermore, an electromagnetic and mechanical re-design of the existing {CERN-}QPR was done with the goal to optimize the measurement range and 
    {the} resolution in~\cite{delPozoRomano:2678067}.}
    \begin{figure}[!tbh]
    \centering
    \includegraphics[width=3.4in]{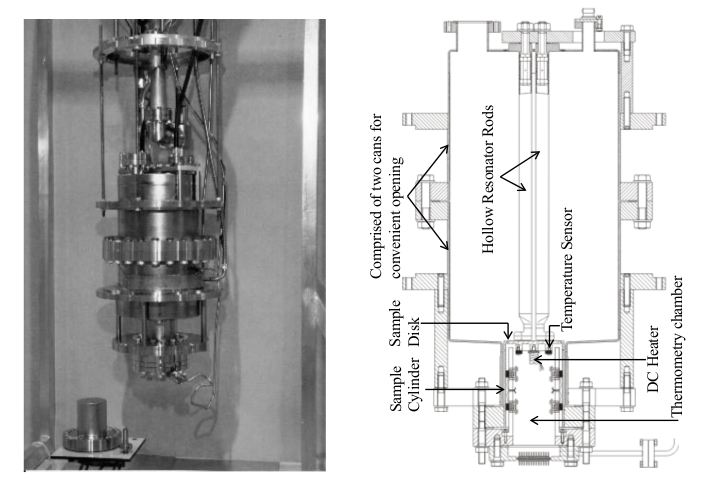}
    \caption{Illustration of the {CERN-}QPR~\cite{Tobias_Junginger2012} (left),  technical drawing~\cite{IEEEhowto:Mahner} (right).
    }
    \label{fig:layout_drawing}
    \end{figure}
    The {\color{teal}HZB-}QPR is used in our research as a case study. The mechanical layout with the technical drawing of the QPR is shown in Figure~\ref{fig:layout_drawing}.
    Its niobium screening cylinder consists of two separate niobium cans. They are electron-beam welded and vacuum-brazed to stainless steel flanges~\cite{Tobias_Junginger2012}. 
    In the center of the first can, four-wire transmission lines are placed, made of niobium rods, which are connected to the upper cover plate of the QPR. 
    These rods are hollow to enable liquid helium to flow inside and maintain the operating condition.
    At the bottom ends, the rods are shorted pairwise in the form of half rings.
    Furthermore, the calorimetry chamber, thermally isolated from the cavity, is mounted at the bottom of the cylinder, below the two loops.
    In this way, the resulting magnetic fields are focused onto the sample, which results in power dissipation, measured by temperature probes inside the calorimetry chamber. In consequence, the surface resistance is investigated using the calorimetric \textit{"RF-DC-compensation"} method~\cite{IEEEhowto:Mahner}. 
    \begin{figure}[!tbh]
    \centering
    \includegraphics[width=3.1in]{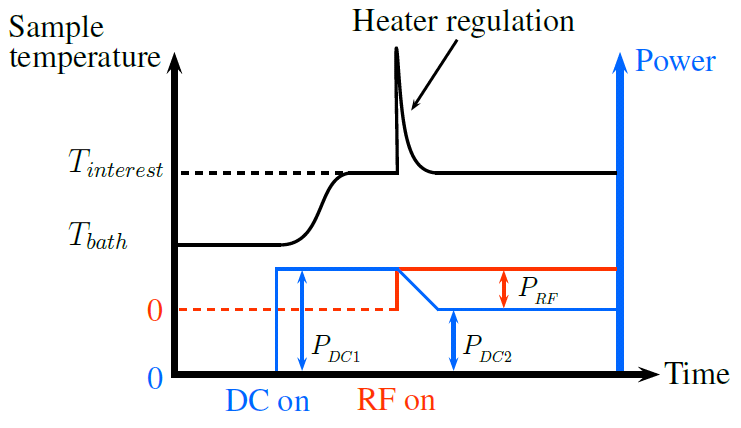}
    \caption{According to the calorimetric method, the surface resistance of a superconducting sample is derived from a DC measurement{~\cite{IEEEhowto:Kleindienst}}} 
    \label{fig:meaus_principle}
    \end{figure}
    
    \subsection{Measurement Principle}
    
   To measure the surface resistance of a sample, the QPR utilizes the \textit{"RF-DC-compensation"} method, which idea is depicted in Figure~\ref{fig:meaus_principle}. First, the sample is heated to a desired temperature of interest $T_{\rm int}$ using the DC heater, which operates in a feedback loop with a proportional–integral–derivative (PID) controller. This allows for determining the heater power $P_{\rm DC1}$ required for temperature stabilization. Next, the RF is turned on, which results in increasing the heat load on the sample. Then, the temperature controller reduces the power in order to reach the thermal equilibrium for the initial temperature $T_{\rm int}$. In steady state, the reduced heater power $P_{\rm DC2}$ is determined and recorded. Hence, the RF dissipated power on the sample surface ${\Omega_{\rm S}}$ is defined by the difference in the DC heater power $\bigtriangleup P_{\rm RF}({\bf p}): = \left[ P_{\rm DC1}({\bf p})-P_{\rm DC2}(\bf p) \right] $ and it is given by   
    \begin{align*}
    \label{power}
        \bigtriangleup P_{\rm RF}({\bf p}) = \frac{1}{2} \int_{\Omega_{\rm S}} R_{\rm S}({\bf p})\, \|{\bf
         H({\bf p})}\|^2 \mbox{d}{\xv},
    \end{align*}
    where ${\bf H}({\bf p})$, $R_{\rm S}({\bf p})$ and ${\bf p}$ denote the magnetic field, the surface resistance and a vector of certain geometrical parameters, respectively. $\| \cdot \|$ denotes the induced norm of complex-valued functions in the space $L^2(\Omega)$. 
    Thereby, under assumptions that $R_{\rm S}({\bf p})$ is independent of ${\bf H}({\bf p})$ and homogeneously distributed across ${\Omega_{\rm S}}$, the surface resistance is defined as
    \begin{equation}
    \label{ssurface_ressi}
        R_{\rm S}({\bf p}) \doteq \frac{\left[ P_{\rm DC1}({\bf p})-P_{\rm DC2}({\bf p}) \right]}{\displaystyle \frac{1}{2} \int_{\Omega_{\rm S}} \|{\bf
         H({\bf p})}\|^2\, \mbox{d}{\xv}},
    \end{equation}
    where the integral term appearing in the denominator is computed numerically as a product of a
    simulation constant $c$ and the stored energy $U$ in the cavity. The latter quantity is measured using a pickup antenna. For details, we refer to
    ~\cite{IEEEhowto:Mahner,Aull}. Consequently, various sources of uncertainties are associated with the \textit{"RF-DC-compensation"} {method}. 
    
    \subsection{Measurement bias due to uncertainties}\label{subsec: meas_bias__uncertain}
     
    Taking this into account, the following sources of uncertainties as well as the systematic measurement error have impact on the measurement resolution and precision~{(further details {are} given in \cite{IEEEhowto:Kleindienst})}:  
    \begin{itemize}
    \renewcommand\labelitemi{--}
      
     \item The bias of the RF power measurement due to the power meter and the cable calibration amounts to $0.2$ [dB] and {$0.1$} [dB], respectively.
      
     \item {The accuracy of the required simulation constant $c$, which relates the denominator of Eq.~\ref{ssurface_ressi} to the stored energy in the cavity.} The associated uncertainty depends (a) on the precision of the numerical computation of the eigenmodes and (b) the geometrical deviations of the physical resonator with respect to the ideal one considered in the simulation. The latter is hard to estimate. For this reason, the standard deviation $5\%$ of $c$ is assumed around a nominal value. 
     
     \item The nonuniform heat distribution of the RF field {on the sample} is negligibly small compared to the other measurement biases.
     
     \item Unwanted heating of the normal-conducting flange{, that is located below the sample cylinder,} results in significant measurement bias for high quality {factor} samples, leading to {overestimation of} residual resistance.
     
     \item Finally, the pulsed measurement results in stochastic uncertainty.
    \end{itemize}
    In particular, the proper functioning of the QPR is negatively influenced by manufacturing imperfections including the roughness of the superconducting surfaces, lack of parallelism of the sample surface and the quadrupole pole shoes as well as insufficient concentric alignment of the coaxial structure~\cite{Keckert2015, putek2019}. 
    
    \subsection{QPR figures of merit}
	
\begin{figure}[!tbh]
    	\centering
    	\includegraphics[width=\columnwidth]{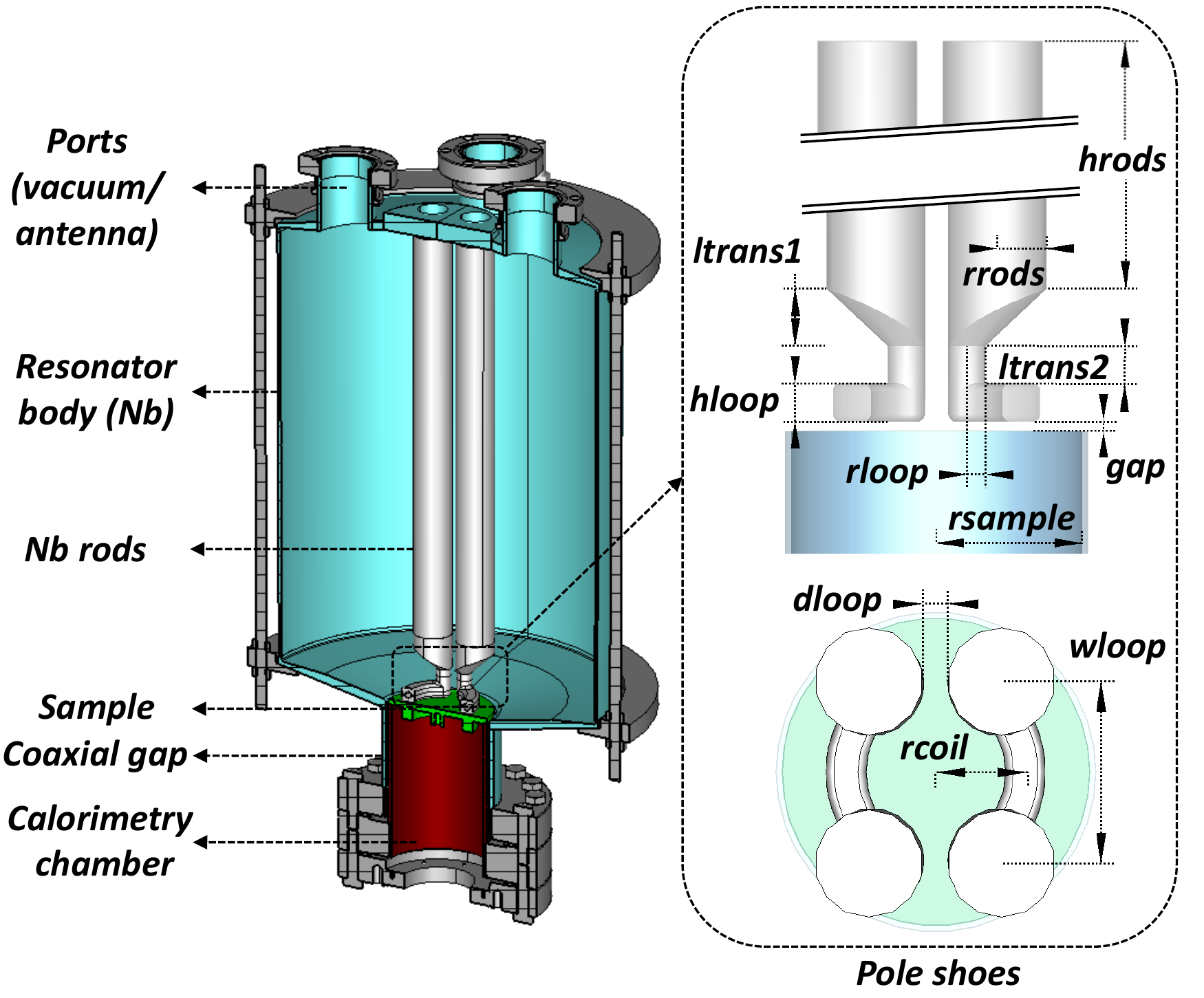}\hfill
    	\caption{Cross section of a {HZB-}QPR (left) and a parameterized model of the pole shoes (right). The nomenclature used in the parameterization of the pole shoes follows~\cite{IEEEhowto:Kleindienst}.} 
    	\label{fig:QPRCrossSection}
    \end{figure}

    Let $D\in \mathbb{R}^3$ denote the computational domain of the QPR and $\Omega \subset D$ be the sub-domain, which is parameterized by variables ${\bf p}=(p_1,\ldots,p_Q)^{\top}$, as shown in Figure~\ref{fig:QPRCrossSection}. Then, to investigate the impact of the uncertain domain $\Omega({\bf p})$ via ${\bf H}:={\bf H}_{\Omega}$ on the performance of the QPR, the following figures of merit are considered~\cite{IEEEhowto:Kleindienst}: 
    \begin{subequations}
    \begin{itemize}
        \renewcommand\labelitemi{--}
        \item Operating modes (frequencies) of the QPR
            \begin{align}
                 \label{f1k}
                 & f_{0,1}({\Omega}) = 0.429 \text{~[GHz]}, \nonumber \\ 
                 & f_{0,2}({\Omega}) = 0.866 \text{~[GHz]}, \\ 
                 & f_{0,3}({\Omega}) = 1.311 \text{~[GHz]}.  \nonumber
            \end{align}
        \item {The denominator of Eq.~(\ref{ssurface_ressi}) normalized to the stored energy 
        {represents} the focus of ${\bf H}$ onto the surface of sample ${\Omega_{\rm S}}$ and is given as}
            \begin{align}
                \label{f2k}
                & f_{1,n}(\Omega, {\bf H}):=\frac{1}{2U}\int_{\Omega_{\rm S}} \|{\bf H}\|^2\,\mathrm{d}{\xv}.
            \end{align}
        {This parameter is also referred to by the symbol $c$.} A higher value of $f_{1,n}({\cdot})$ implies an improvement of the measurement resolution and, consequently, an increase of the measurement signal.
        \item {The homogeneity of the magnetic field distribution on the surface of {the} sample is represented by the following dimensionless quantity}
            \begin{align}
                \label{f3k}
                 & 	f_{2,n}(\Omega, {\bf H}):= \frac{\displaystyle \int_{\Omega_{\rm S}}^{}\|{\bf H}\|^2\,\mathrm{d}{\bf x}}{|\Omega_{\rm S}|\,\displaystyle{\max_{{\bf x} \in \Omega_{\rm S} }(\|{\bf H}\|^2)}},
            \end{align}
        which maximizes the measurement signal through the increase of the dissipated power at given magnetic peak field {$\displaystyle{\max_{{\bf x} \in \Omega_{\rm S} }(\|{\bf H}\|^2)}$ on ${\Omega_{\rm S}}$.}
        
        \item {The penetration of the magnetic field into the coaxial gap around the sample cylinder might lead to heating up of the normal-conducting flanges and subsequently gives rise to measurement bias. The following dimensionless parameter quantifies the penetration of the magnetic field into the coaxial gap at a given loss on the surface of sample}
        \begin{align}
            \label{f6k}
        	f_{3,n}(\Omega, {\bf H}):=\frac{\displaystyle \int_{\Omega_{\rm S}}^{}\|{\bf H}\|^2\,\mathrm{d}{\bf x}}{ \displaystyle \int_{\Omega_{\rm F}}^{}\|{\bf H}\|^2\,\mathrm{d}{\bf x}},
        \end{align}
        {where ${\Omega_{\rm F}}$ is related to the region of the flange. In this paper the denominator of Eq.~(\ref{f6k}) is evaluated in the coaxial gap at \unit[7]{cm} below the surface of sample.}
        
        \item  {The peak surface magnetic field inside the QPR is typically located on the rods. This can limit the maximum attainable field on the sample due to the magnetic break down limit of the superconducting materials. Thus, the following dimensionless parameter should be maximized} 
        \begin{align}
            \label{f4k}
             & f_{4,n}(\Omega, {\bf H}): = {\frac{\displaystyle{\max_{{\bf x} \in \Omega_{\rm S} }(\|{\bf H}\|)}}{\displaystyle{\max_{{\bf x} \in \Omega_{\rm R} }(\|{\bf H}\|)}}},
        \end{align}
        {where $\displaystyle{\max_{{\bf x} \in \Omega_{\rm S} }(\|{\bf H}\|)}$ and $\displaystyle{\max_{{\bf x} \in \Omega_{\rm R} }(\|{\bf H}\|)}$ are the peak magnetic field on the sample ${\Omega_{\rm S}}$ and on the rods ${\Omega_{\rm R}}$, respectively.}

         \item {The limitations caused by high surface electric field on the rods, e.g., field emission, could be lowered by maximizing} 
        \begin{align}
            \label{f5k}
             & f_{5,n}(\Omega, {\bf H}):={\frac{\mu_0 \displaystyle{\max_{{\bf x} \in \Omega_{\rm S} }(\|{\bf H}\|)}}{\displaystyle{\max_{{\bf x} \in \Omega_{\rm R}}(\|{\bf E}\|)}}},
        \end{align}
        {where $\displaystyle{\max_{{\bf x} \in \Omega_{\rm R}}(\|{\bf E}\|)}$ denotes the maximum electric field on the rods and $\mu_0$ is the magnetic permeability of vacuum.}
        
    \end{itemize}
    {The subscripts $n=1,2,3$ indicate the particular operating frequency and $\|\cdot\|$ is the induced norm of complex-valued functions.}
    
    \end{subequations}

	It can be summarized that the associated uncertainties have significant impact not only on the outcome of the measurement methodology, but also on the stability and operational conditions of the QPR. Therefore, in this work, the uncertainty propagation in the three-dimensional (3D) model of the QPR is investigated. Here, special emphasis is laid on the influence of the
    geometrical parameters on reliable operation of the QPR under working conditions.
    {Thus, in Sections~\ref{sec:StochforwProb} and~\ref{sec:moso} the problem listed in subsection~\ref{subsec: meas_bias__uncertain} as the second item of uncertainty source is mainly addressed. The third item (nonuniform heat distribution) is also included by considering Eq.~(\ref{f3k}) as one of the objective functions.}

    \section{Stochastic forward problem}
    \label{sec:StochforwProb}

    This section briefly discusses a 3D model of the QPR that is governed by the Maxwell's eigenproblem (MEP). Next, the uncertainty of geometric parameters is introduced into the MEP model and described in a probabilistic framework. This allows {for} mimicking manufacturing imperfections appearing in the industrial process.

	\subsection{Problem setup in deterministic settings}

	Let ${\bf p} = (p_1, \ldots,p_Q )^{\top}\in \Pi \subset \mathbb{R}^Q$ denote a vector of geometrical parameters, for example, these variables that are depicted in Figure~\ref{fig:QPRCrossSection}. 
	Furthermore, denote by $D \subset \mathbb{R}^{d},\, d = 3$, a bounded and simply connected physical domain with 
	sufficiently smooth boundary $\partial D$ representing the QPR structure shown in Figure~\ref{fig:QPRCrossSection}. 
	{Next, suppose $\epsilon_r$ and $\mu_r$ stand for the relative electric permittivity and the relative magnetic permeability, which are linear functions of space.} 
	Then, the associated eigenpairs $({\bf E}({\bf p}), \lambda({\bf p}))$ for each mode given in terms of the phasor of the electric field eigenvector ${\bf E}$ and the eigenfrequency $\lambda = \frac{{\omega^2}}{{c_0^2}}$ satisfy the MEP for electric fields in the time-harmonic regime     
	\begingroup\makeatletter\def\f@size{9.5}\check@mathfonts
	\begin{align}
	    \label{eq:strongf}
        & - \nabla  \times \left( \frac{1}{\mu_r}\, \nabla \times {\bf E}({\xv, {\bf p}})\right) + \lambda({\bf p})\,  \epsilon_r\, {\bf E}({\xv, {\bf p}})= 0,&\text{in } D, \nonumber \\  
        &{\bf n} \times {\bf E}({\xv, {\bf p}})  = 0, & \text{on } \partial D_{\rm P}, \nonumber \\ 
        &{\bf n} \times \left( \frac{1}{\mu_r}\, \nabla \times {\bf E}({\xv, {\bf p}})\right) = 0, & \text{on } \partial D_{\rm N} 
     \end{align}
     \endgroup
	with the angular frequency $\omega =2\pi f$ and the speed of light $c_0$. Here, $\partial D_{\rm P}$ denotes the portion of the boundary with the perfect electrical conductor (PEC) condition, while $\partial D_{\rm N}$ is the portion of boundary with Neumann condition and ${\bf n}$ the outward unit normal to the boundary, where $\partial D = \partial D_{\rm P} \cup \partial D_{\rm N}$. 
	
	Next, the uncertain parameters ${\bf p}$ need to be specified.
	
	\subsection{Probabilistic framework for uncertainty }\label{subsec:pf}  
	
	The probabilistic framework~\cite{IEEEhowto:Xiu07} is used for modeling geometrical uncertainties as $Q$-variate random vectors with independent components. It is assumed that they are defined in the probabilistic space $\left(\mathcal{A},  \mathcal{F}, P \right)$, where $\mathcal{A}$ is a sample space, $\mathcal{F}$ denotes a sigma algebra, and $P : \mathcal{F} \rightarrow \left[0, 1\right]$ refers to a probability measure. 
	Furthermore, we denote by $\rho_{q}:~\Gamma_{q} \rightarrow \mathbb{R}^{+}$ the probability density function (PDF) of the random variable ${p_q(\xi)},~\xi \in \mathcal{A}$ and by $\Gamma_{q}\equiv p_{q}(\mathcal{A})\in \mathbb{R}$ the image of $p_{q}$ with its support $\Gamma = \prod_{q=1}^{Q} \Gamma_{q} \subset \mathbb{R}^Q$, for $q\,=\, 1, \ldots, Q$. Then, a joint PDF of the random vector ${\bf p}\left({\bm \xi}\right)$ is assumed to exist and is given by
    \begin{align}
        \label{eq:dense}
        \rho\left({\bf p}\right)=\prod_{q=1}^{Q} \rho_{q}\left(p_{q}\right), 
    \end{align}
    where the dependence of ${\bm \xi}$ has been suppressed.
    Finally, the modified random vector of geometrical parameters is given by 
    \begin{align}
    \label{eq:rand_param}
        {\bf p}(\bm \xi) = (p_1(\xi_1),\ldots,p_Q(\xi_Q))^{\top} : \Gamma \rightarrow \Pi \subset \mathbb{R}^Q.    
    \end{align}
    Consequently, the application of the probabilistic framework for modeling uncertainty allows one to conduct numerical formulations in the finite dimensional random space $\left(\Gamma, \mathcal{B}^Q, \rho{\rm d}{\bf p} \right)$ with $\mathcal{B}^Q$ being the $Q$-dimensional Borel space~\citep{Okse03}. 
    
    \subsection{Stochastic Maxwell's Eigenproblem}
    
    Consider now the random complex function ${\bf u} : \Gamma \rightarrow \mathbb{C}^d$, for which the probabilistic Hilbert space on the complex field is introduced $L^2(\mathrm{\Gamma)} = \{ {\bf u}({\bf p}) : \mathbb{E} [\, \|{\bf u}({\bf p})\|^2\,] < \infty \}$, see, e.g.,~\citep{Malliavin}. Then, the expected value of ${\bf u}({\bf p})$ is defined as
    \begin{align} \label{mean}
    \mathbb{E} [\, {\bf u}({\bf p})\, ] := 
    \int_{\Gamma} {\bf u}({\bf p})\, \rho \, \mbox{d} {\bf p}.  
    \end{align}
   Likewise, an inner product for two random functions ${\bf u}({\bf p}),{\bf v}({\bf p})) : \Gamma \rightarrow \mathbb{C}^d$ is given by
    \begin{align} \label{innerproduct}
        \big(\, {\bf u}({\bf p}) , {\bf v}({\bf p}) \,\big)_{L^2_{\rho}(\Gamma)}: =  \int_{\Gamma} {\bf u}({\bf p}) \cdot \overline{{\bf v}({\bf p}})\, \rho \; \mbox{d} {\bf p}, 
    \end{align}
    where $\overline{\bf v({\bf p})}$ denotes the complex conjugate. Moreover, based on the definition \eqref{mean} the variance of a random complex function ${\bf u}({\bf p}) \in L^2 (\Gamma)$ reads as
    \begin{equation} \label{variance}
        {\rm Var} [ {\bf u}({\bf p}) ] := \mathbb{E} [ \|\,{\bf u}({\bf p})\,\|^2 ] - \|\,\mathbb{E} [ {\bf u}({\bf p}) ]\,\|^2, 
    \end{equation}
    which is always real and positive.
    
    Finally, the weak formulation of the MEP with random input parameters ${\bf p}$ is given as follows:\\ Find ${\bf E} \in V_{\rho}$ such that
    \begingroup\makeatletter\def\f@size{9.0}\check@mathfonts
    \begin{align}
    \label{eq:st_weak}
    \mathbb{E}\left[ \int_{D}\,{\frac{1}{\mu_r}\, (\nabla \times {\bf E})\cdot (\nabla \times {\bf v})}\,\mbox{d}{\xv}\right] = \mathbb{E}\left[ \int_{D} \lambda\,  \epsilon_r\, {\bf E} \cdot {\bf v}  \mbox{d}{\xv}\, \right]
    \end{align}
    \endgroup
    is satisfied for all ${\bf v} \in V_{\rho}$ with $V_{\rho}$ defined as the tensor product $V_{\rho} = H^1_0(D) \otimes L^2_{\rho}(\Gamma)$, where $H^1_0$ is the Sobolev space of the complex-valued functions with first order weak derivatives and the $0$ subscript refers to vanishing tangential component of ${\bf E}$ on $\partial D_{\rm P}$, see, e.g.,~\cite{Monk2003}. 
    
    The variational formulation of \eqref{eq:st_weak} involves expectations of the weak form, formulated in the physical space, which can be solved using, {e.g.,} {the finite elements method (FEM)} {or the Finite Integration Technique (FIT)~\cite{cst2018}}. For the solution of~\eqref{eq:st_weak}, the non-intrusive method called PC-SCM~\cite{Xiu05,IEEEhowto:Xiu07} (also known as pseudo-spectral approach) is preferable. 
    This method is outlined below in the next section.

	\section{Pseudo-spectral approach}\label{sec:PseuSpectrAppr}
	In this section, the mathematical framework of the PC-based SCM~\cite{IEEEhowto:Xiu07} is shortly presented. For this reason, the mathematical bases of the PC expansion will be briefly introduced. 
	
	\subsection{Polynomial chaos expansion}

    The homogeneous PC was introduced in~\citep{Wien38}. It employs the Hermite orthogonal polynomials in terms of Gaussian random variables to provide the spectral expansion of the stochastic processes. This idea, furthermore, has been revisited in~\cite{Ghanem} and applied to the field of engineering. More recently, a broader framework, the so-called generalized PC, has been developed by~\cite{Xiu02}. It is based on the Wiener-Askey scheme and allows for representing more efficiently non-Gaussian processes as well. According to the theory of~\cite{Wien38, Cameron1947, Xiu05} any second-order random function $y \in L^2(\Gamma)$ can be represented by a weighted sum of polynomials $\Phi(\bf p)$, which are dependent on random variables ${\bf p}$ of the known PDF $\rho({\bf p})$.
    
    Let ${\bm i}$ be a multi-index ${\bm i} = (i_1, \ldots, i_Q) \in \mathcal{I}_{Q,P}$, where $P$ denotes the polynomial order. Next, denote by $\mathcal{I}_{Q,P}$ the set of multi-indices, which is defined as
    \begin{align*}
        \mathcal{I}_{Q,P} = \{{\bm i} = (i_1, \ldots, i_Q) \in \mathbb{N}^Q_0 : |{\bm i}| \leq P\}.
    \end{align*}
    where $|\,\cdot\,| := i_1 + \ldots + i_Q$ is the $l_1$ norm. %
    Then, given a square-integrable, random complex function with finite variance $y \in L^2(\Gamma)$, a truncated PC expansion is introduced~\cite{Xiubook, Xiu05, Xiu02, IEEEhowto:Xiu07, Pulch2012}  
    \begin{align}
        \label{eq:trun}
         y({\bf p}) \doteq \sum_{{\bm i}\in\mathcal{I}_{Q,P}}\, \widetilde{y}_{\bm i}\, \Phi_{\bm i}({\bf p}), \quad \widetilde{y}_{\bm i} \in \mathbb{\color{black}C}.
    \end{align}
    Here, $\widetilde{y}_{\bm i}$ are a priori unknown coefficient functions to be determined, while the multivariate PC basis functions $\Phi_{\bm i}({\bf p})$ are generated from
    \begin{align}
        \Phi_{\bm i}({\bf p}) = \prod_{k=1}^Q\,\Phi_{i_k}(p_k), \quad {\bm i} \in \mathcal{I}_{Q,P},  
    \end{align}
    where $\Phi_{i_k}(p_k)$ are univariate polynomials of degree $i_k \in \mathbb{N}_0$, which are orthogonal with respect to $\rho_k(\bf p)$. A popular choice for the functions $\Phi_{i_k}$ are orthonormal polynomials\footnote{In the case of an orthogonal system of basis polynomials a normalization can be performed easily, see, e.g., \citep{Xiubook}}. Let $\Phi_{\bf 0}:=1$. Thus, when using \eqref{innerproduct} it follows that
    \begin{align}
    \label{eq:ortho}
    \big(\, \Phi_{i_q} , \Phi_{j_q} \,\big)_{L^2_{\rho}(\Gamma)}&: = \int_{\Gamma} {\Phi_{i_q}}({p}_q) {\Phi_{j_q}}({p}_q)\, \rho ({p}_q) \; \mbox{d} {{p}_q}\\ \nonumber
    & = \left\{ \begin{array}{ll}
    0 & \;\mbox{ for}\;\; i_q \neq j_q \\
    1 & \;\mbox{ for}\;\; i_q = j_q. \\
    \end{array} \right. 
    \end{align}
    Certainly, the condition \eqref{eq:ortho} and the independence of $p_q$ imply the orthonormality of $\Phi_{\bm i}({\bf p})$.
    The number $K$ of PC basis functions of total order $P$ in dimension $Q$ is given by
    $$ K = |\mathcal{I}_{Q,P}| = \frac{(P+Q)!}{P!Q!}.$$
    
    The truncated PC expansion in \eqref{eq:trun} converges in the mean-square sense under following conditions~\cite{Xiubook}:
   \begin{itemize}
   \renewcommand\labelitemi{--}
   \item $y(\bf p)$ has finite variance
   \item the coefficients $\widetilde{y}_{\bm i}$ are calculated from the projection equation
    \begin{align}
    \label{eq:proj}
        \widetilde{y}_{\bm i}: = \big(\, y({\bf p}) , \Phi_{\bm i}({\bf p}) \,\big)_{L^2_{\rho}(\Gamma)}= {\mathbb{E}[y({\bf p})\Phi_{\bm i}({\bf p})]}.    
    \end{align}
   \end{itemize}
   In general, Eq.~\eqref{eq:proj} gives rise to two main methods, which explore the projection equation in a different way. The first one is the spectral Galerkin method~\cite{Ghanem}, which belongs to the intrusive techniques and applies Eq.~\eqref{eq:proj} and, consequently, Eq.~\eqref{eq:ortho} to project governing equations. As a result, a dedicated solver needs to be used to solve the resulting huge system of equations due to the spectral expansion. The second, non-intrusive technique, i.e., the pseudo-spectral approach~\cite{IEEEhowto:Xiu07}, which, likewise the Monte Carlo methods, allows for reusing existing deterministic solvers but in a much more efficient way, in case of smooth models. In this work, the focus is laid on the latter approach.
   
   \subsection{Uncertainty Quantification via PC expansion}
   
   The pseudo-spectral approach~\cite{IEEEhowto:Xiu07} applies the projection equation only to output quantities of interest. As a typical non-intrusive method, it only requires repetitive simulations of a deterministic model at quadrature points ${\bf p}^{(k)} \in \Gamma, k=1,\ldots, K$. Then, the discrete projection of given solutions $y({\bf p}^k)$ on the basis polynomials $\Phi_{\bm i}$ by using the multi-dimensional quadrature with associated weights $w_k$
   \begin{align}
    \label{eq:disc_proj}
        \widetilde{y}_{\bm i}
        \doteq  \sum_{k=1}^{K} w_k\, y\left({\bf p}^{(k)}\right)\,\Phi_{\bm i}\left({\bf p}^{(k)}\right),
    \end{align}
    yields the approximation of probabilistic integrals~\eqref{eq:proj}. The effectiveness of this approach strongly depends on the choice of quadrature nodes. If not carefully chosen the straightforward application of the tensor product of a one-dimensional Gauss interpolation formula might become computationally expensive. Thus, to overcome the so-called curse of dimensionality problem, either the Smolyak algorithm~\cite{Smolyak1963Q,Petras2003} or the effective Stroud~\cite{Stroud,Xiu:2008} formulas can be applied. 
    
    In general, the Stroud integration rules yield uniform, beta or normally distributed points, which are weighted by $\frac{1}{N}$ with $N$ denoting the number of points. Specifically, in this work, normally distributed points generated by the Stroud-3 formula are considered. This choice is motivated by the physics of the analyzed application, i.e., due to a lack of statistical data it is assumed that the geometrical tolerance of the QPR design caused by the {manufacturing} process is normally distributed. Correspondingly, in a model with $Q$ uncertain parameters, only $2Q$ quadrature points are required{~\cite{Benner_2015}}. Now, consider, for example, the $j$-th component of the normally distributed points around the mean  ${\overline{\bf p}_j}$ with the standard deviation ${\bm \sigma}_j$,~\cite{Benner_2015,Bagci09}
    \begin{align*}
        {p}_j^{i} = {\overline{\bf p}_j} + {\bm \sigma}_j \cdot { z}_q^{j}
    \end{align*}
   with $i$ given as $i=2r-1,\, i=2r$, respectively, for $r = 1,\ldots,\lfloor Q/2\rfloor$
   \begin{align}
    \label{eq:nodes}
       &z_j^{2r-1} = \sqrt{2}\cos{\left(\frac{(2r-1)j\pi}{Q}\right)},\nonumber\\
       \\
       &z_j^{2r} = \sqrt{2}\sin{\left(\frac{(2r-1)j\pi}{Q}\right)}. \nonumber
   \end{align}
    Here, if $Q$ is odd, then $z_j^Q = (-1)^j$, while an operator $\lfloor Q/2\rfloor$ returns the largest natural number smaller or equal than $Q/2$. Though, the Stroud formulas are very effective because they yield a very small number of quadrature points, they have also a fixed accuracy. 
    
    \subsection{Statistical information \& sensitivity analysis}
    
    Due to the orthonormality \eqref{eq:ortho} of the polynomial basis, once the PC expansion \eqref{eq:trun} is found, all statistical information can be retrieved. In particular, the expected value and the variance are given by 
    \begingroup\makeatletter\def\f@size{9.5}\check@mathfonts
    \begin{equation}
    \label{eq:stat_moments}
    \mathbb{E}_{} \left[ {{ y}\left({{ {\bf p}}} \right)} \right]\;\doteq\;{ { 
    \tilde{y}}}_{\bf 0} , \quad \mbox{Var}\left[ {{y}\left({{ {\bm p}}} 
    \right)} \right]\;\doteq\sum\limits_{\substack{{{\bm i}\in\mathcal{I}_{Q,P}}\\{{\bm i}} \neq {\bf 0}}} {\left| {{ {\tilde{y}}}_{{\bm i}} } 
    \right|^{2}}
    \end{equation}
    \endgroup
    using $\Phi_{\bf 0} = 1$. Based on \eqref{eq:trun}, also other quantities such as the local sensitivity, the variance-based global sensitivity, the approximation of the PDF, and of the cumulative PDF can directly be evaluated, see, e.g., \cite{Xiubook}.
    
    For example, the local sensitivity (a partial derivative), i.e., the $p_q$-th mean sensitivity is obtained by integrating over the whole parameter space and it is given by~\cite{Xiu09fastnumerical}
    \begin{equation}
    \label{mean_grad_sens}
    \mathbb{E}\bigg{[}\frac{\partial y}{\partial p_q}\bigg{]} 
    \doteq\sum\limits_{\substack{{{\bm i}\in\mathcal{I}_{Q,P}}\\{{\bm i}} \neq {\bf 0}}}\left( {{ {\tilde{y}}}_{{\bm i}} } \int \frac{\partial  \Phi_{\bm i}({\bf p})}{\partial p_q } \rho \mbox{d} {\bf p} \right)
    \end{equation}
    for $q=1,\ldots, Q.$ On the contrary, the global sensitivity approach does not specify any additional condition as $p_q = {\overline{p}_q}$. Instead, it considers only a model, e.g., \eqref{eq:trun} and analogous decomposition to the ANOVA~\cite{IEEEhowto:sudret2008} {(the ANalysis of VAriance)} is conducted to find the contribution of particular random variables to the total variance. For this reason, it should be regarded as a more reliable tool, especially in the case of modeling and optimization processes.
	\begin{figure}[h] %
    	\centering
    	\includegraphics[width=\columnwidth]{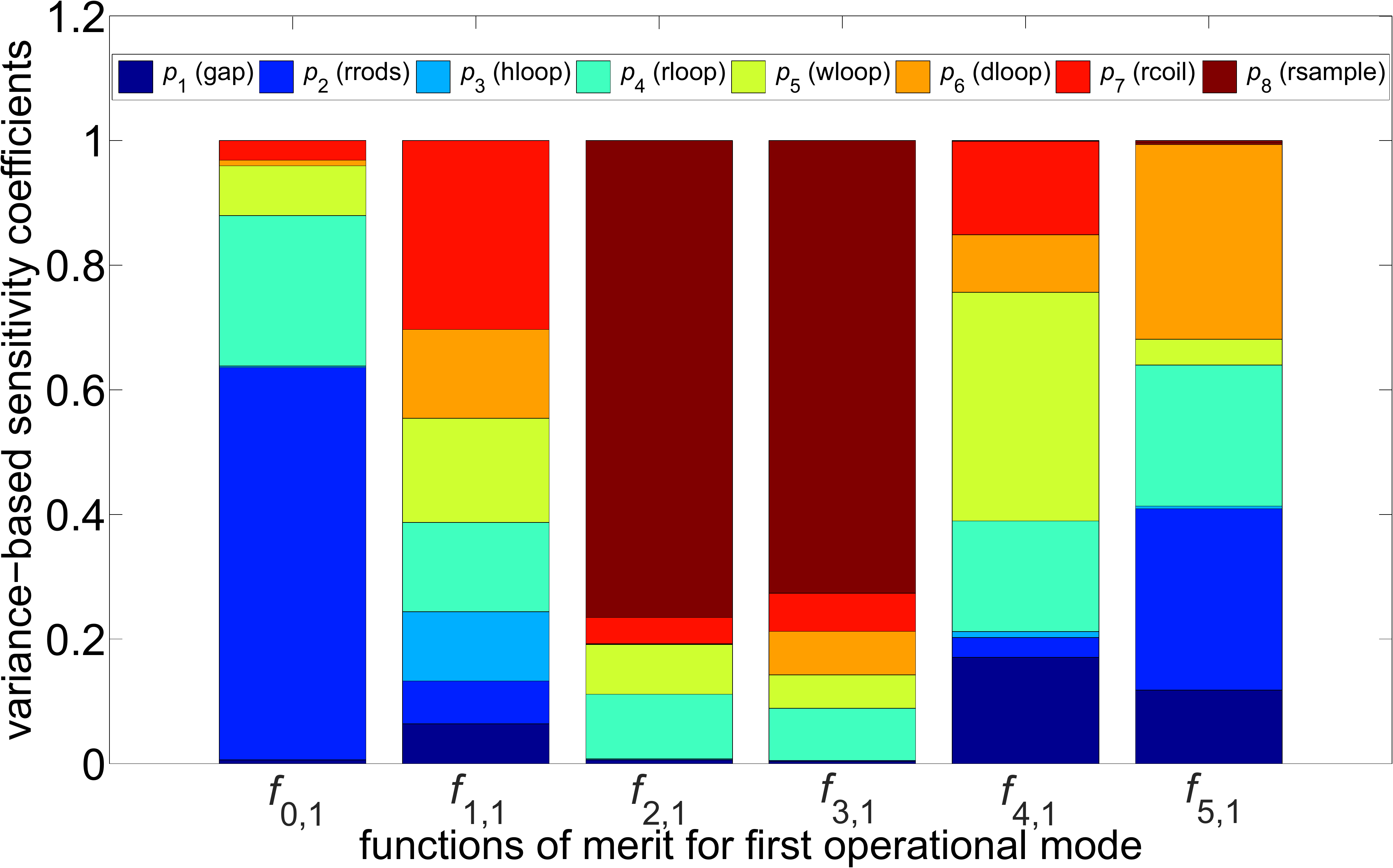}\hfill
    	\caption{Result of the global sensitivity analysis for $f_{\cdot,1}({\bf p})$.}
    	\label{fig:f01}
    \end{figure}
	\begin{figure}[h]
    	\centering
    	\includegraphics[width=\columnwidth]{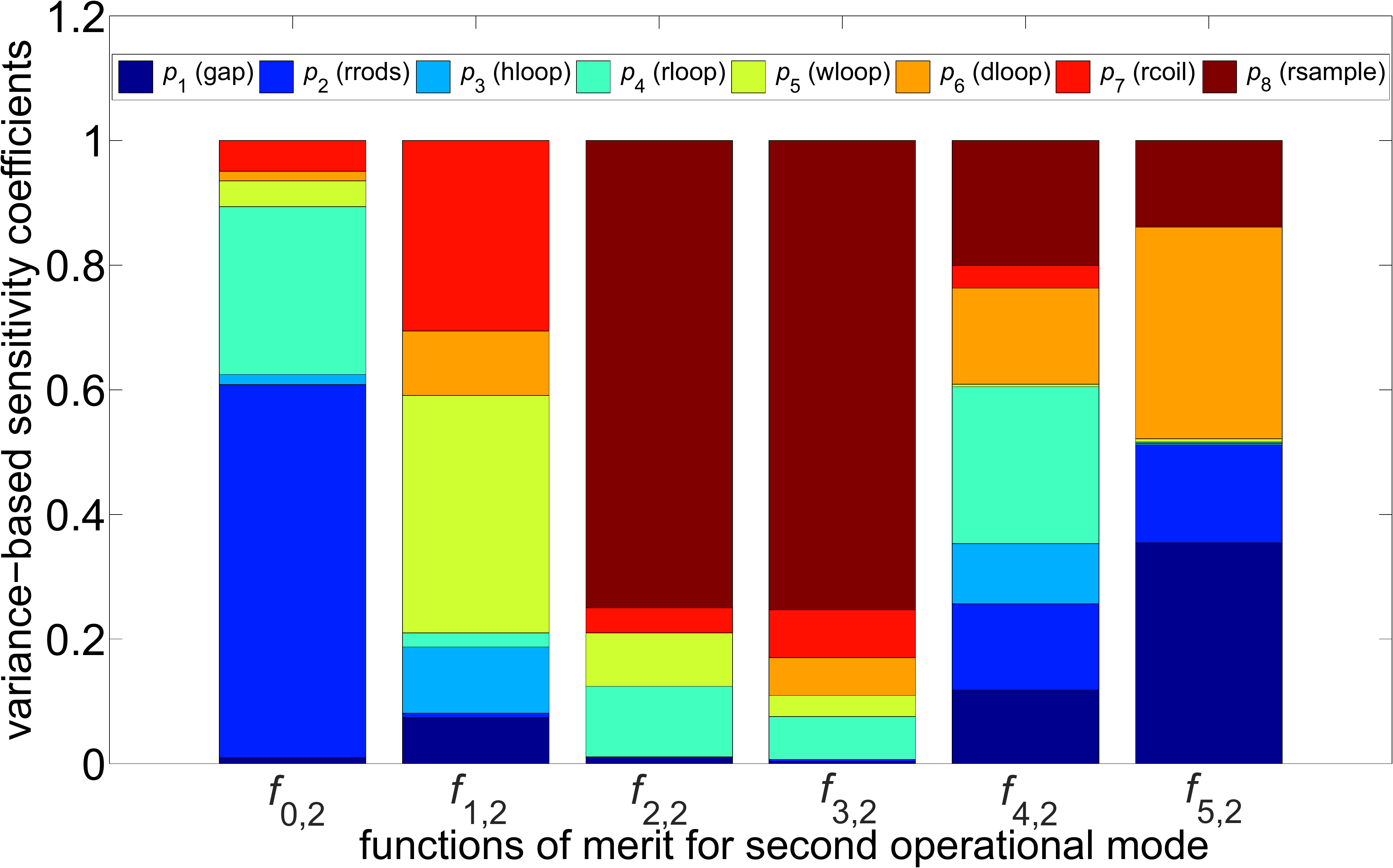}\hfill
    	\caption{Result of the global sensitivity analysis for $f_{\cdot,2}({\bf p})$.}
    	\label{fig:f02}
    \end{figure}
	\begin{figure}[h]
    	\centering
    	\includegraphics[width=\columnwidth]{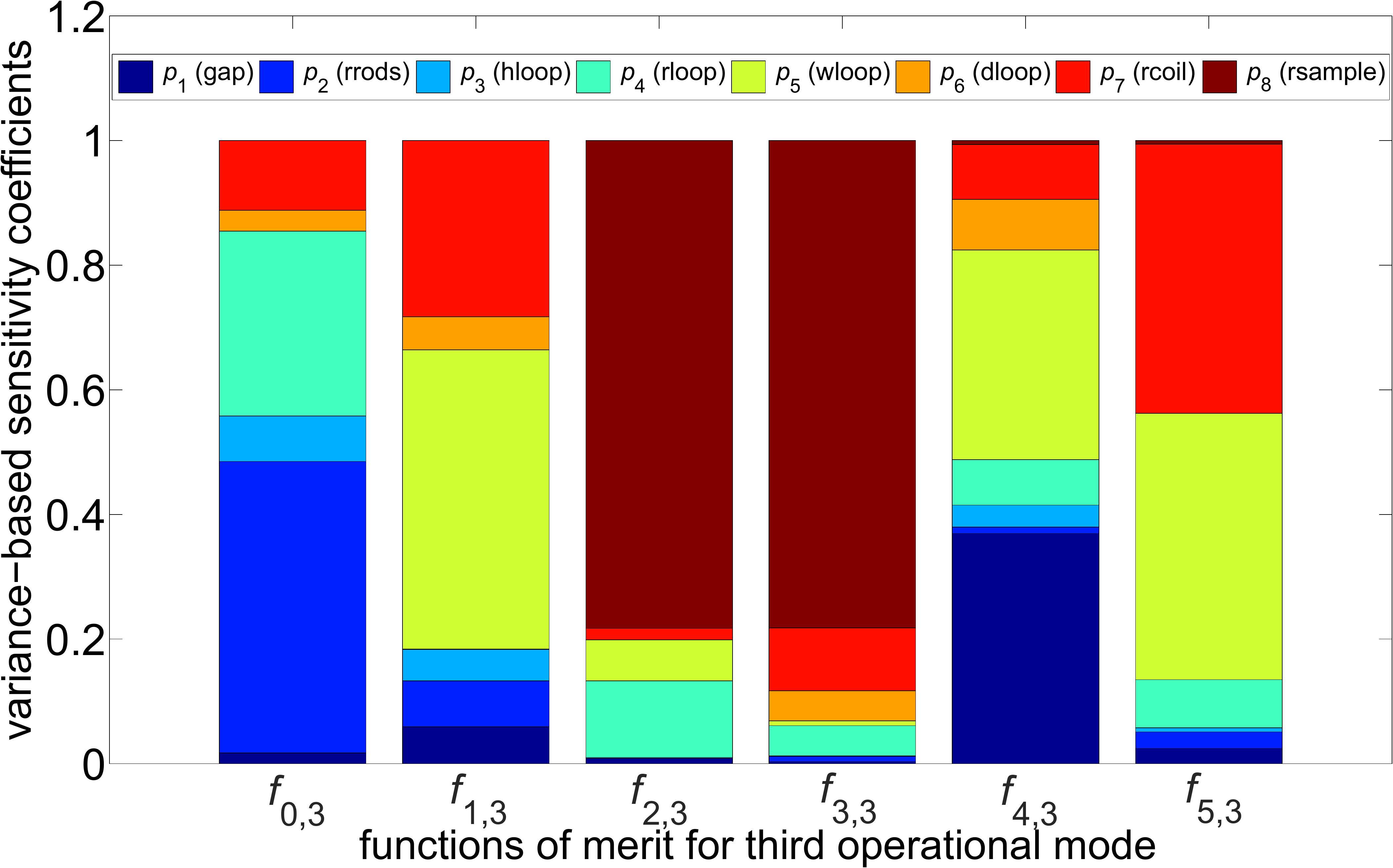}\hfill
    	\caption{Result of the global sensitivity analysis for $f_{{\cdot},3}({\bf p})$.}
    	\label{fig:f03}
    \end{figure}

    The Sobol decomposition of \eqref{eq:trun} yields the (first-order) variance-based sensitivity (VBS) coefficients~\cite{IEEEhowto:sobol} 
    \begin{align}
    \label{glob_sen}
     {\bf S} = (s)_{l,{q}}\in \mathbb{R}^{L,Q}, \quad s_{l,{\color{black}q}} = \frac{1}{\mbox{Var}(y_{l}({\bf p}))}\,\sum_{{\bm i} \in \mathit{I}_{q}}\,|\tilde{y}_{l,{\bm i}}|^2 
     \end{align}
    with sets $\mathit{I}_{q} = \{{\bm i}\in \mathbb{N}_0: i_q > 0, i_{m \neq q}=0\}$ and the total variance of particular objective functions, $l=1,\ldots,L$, denoted by $\mbox{Var}(y_l)$. In order to calculate ${s_{l.q}}$, all random inputs except ${p_q}$ are fixed. Thus a mixed effect, that is, the interactions between $p_q$ and other random variables is neglected here. The upper and lower bounds of $s_{l,q}$ are given by $0\leq s_{l,q}\leq 1$. A value close to $1$ denotes a large contribution to the variance, while a small contribution is determined by a value close to $0$. The total effect, that is, the fractional contribution to the total variation of $y(\bf p)$ due to parameter ${p_q}$, when considering all other model parameters can also be analyzed~\cite{IEEEhowto:sudret2008}. 
    \begin{table}[h]
       \centering
       \caption{Means $\overline{p}_q$ and std. dev. $\sigma_q$ of random inputs }
       \begin{tabular}{|l|r|r|}
           \hline
           \textbf{Name}  & \textbf{$\overline{p}_{q} $} & \textbf{$\sigma_q$} \\
           \hline
              $p_1$~{(gap)}           & {0.54} [mm]     & {0.027} [mm]    \\ 
              $p_2$~{(rrods)}         & {13.40} [mm]    & {0.67} [mm]    \\ 
              $p_3$~{(hloop)}         & 9.50 [mm]     & 0.475 [mm]    \\ 
              $p_4$~{(rloop)}         & 5.00 [mm]     & 0.25 [mm]     \\
              $p_5$~{(wloop)}         & 40.00 [mm]    & 2.000 [mm]     \\
              $p_6$~{(dloop)}         & {6.00} [mm]     & {0.300} [mm]     \\
              $p_7$~{(rcoil)}         & 22.48 [mm]    & 1.124 [mm]     \\
              $p_8$~{(rsample)}       & 38.50 [mm]    & 1.929 [mm]     \\
           \hline
       \end{tabular}
       \label{l2ea4-t0}
    \end{table}
   
    {For example, Figures~\ref{fig:f01}--\ref{fig:f03} present, for all operating modes, the results for VBS analysis of the {quantities of interest} with respect to the Gaussian design parameters, listed in Table~\ref{l2ea4-t0}.
    In particular, it can be observed that deviations of $p_2~{\rm (rrods)}$ and $p_3~{\rm (hloop)}$, have the greatest influence on the operating frequencies, given by Eq.~\eqref{f1k}. The geometric parameters such as $p_5~{\rm (wloop)}$, $p_6~{\rm (dloop)}$ and  $p_7~{\rm (rcoil)}$, in turn, have significantly larger contributions to the the focusing strength, expressed by Eq.~\eqref{f2k}. Correspondingly, the variations of $p_4~{\rm (rloop)}$, $p_8~{\rm (rsample)}$ can be identified to have a large impact on both the homogeneity and dimensionless factors, defined by Eqs.~\eqref{f3k} and~\eqref{f4k}, respectively. In the end, this analysis can guide new designs but also improve the existing QPR configurations. In contrast, the VBS decomposition provided for the magnetic {peak}
    values, that is, Eqs.~\eqref{f5k} and~\eqref{f6k} show the significant differences with respect to design parameters within the range of operating modes.}

\begin{figure}[h] 
    	\centering
    	\includegraphics[width=6.0cm]{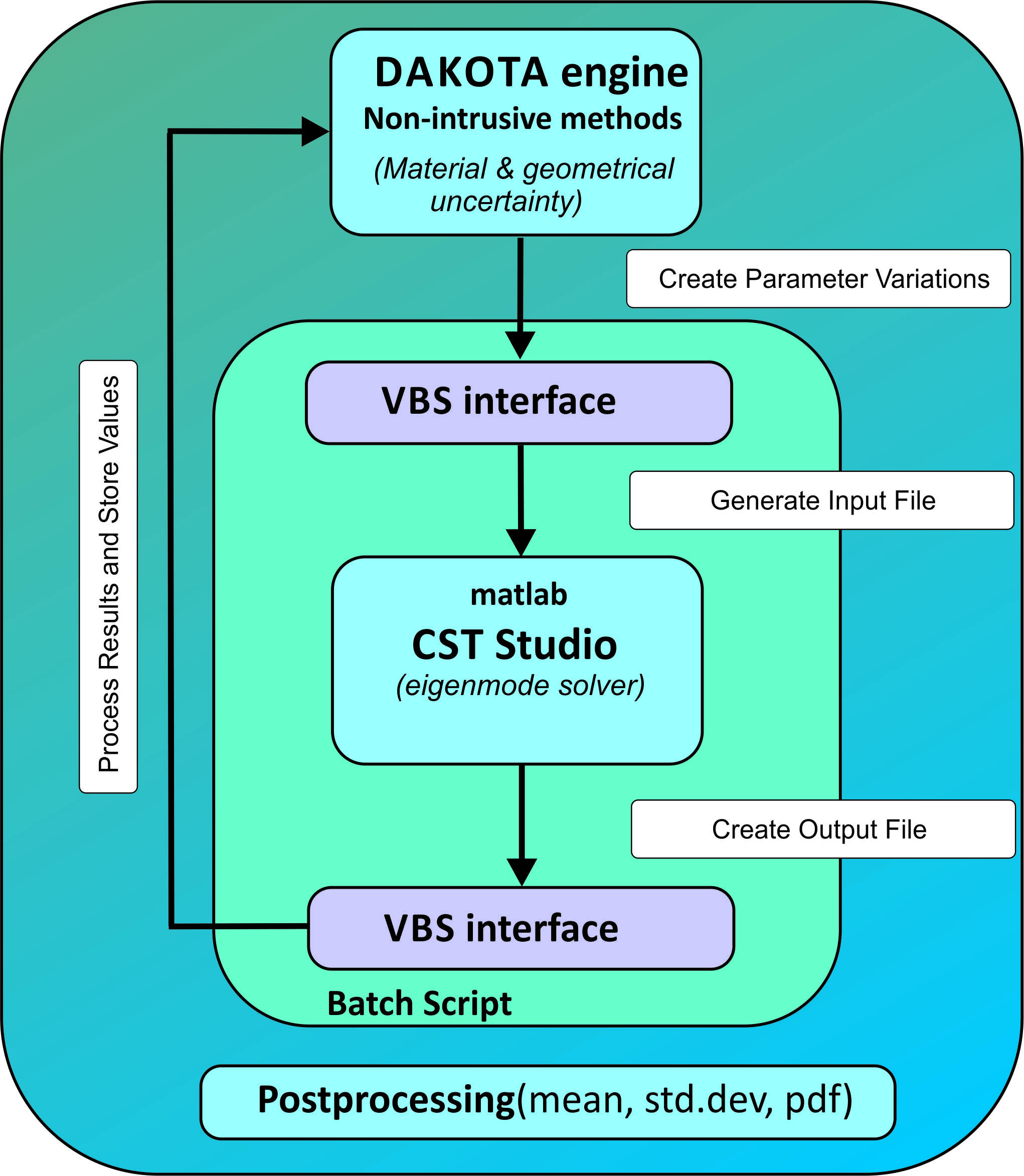}
    	\caption{Algorithm for the PC-SCM employing
    CST STUDIO SUITE\textsuperscript{\textregistered}~\cite{cst2018} as ’black-box’ simulation engine. In the flow indicated above Dakota~\cite{dakota2018} has been exploited.}     	\label{fig:flow}
    \end{figure}

    The flow of the algorithm of the pseudo-spectral method has been shown in Figure~\ref{fig:flow} and, additionally, described in the pseudo-code as Algorithm~\eqref{alg:psauq} in order to allow better understanding.

    In order to find a robust design of the QPR, the UQ analysis needs to be incorporated into the optimization flow. Section~\ref{sec:moso} is devoted to the parametric multi-objective shape optimization under uncertainties. 

    \begin{algorithm}[H]
    \caption{Pseudo-spectral approach for UQ} \label{alg:psauq}
    \begin{algorithmic}[1]
        
        \State Initialization :
        \State \quad -- ${\Pi} = \left( {p}_1,\dots, {p}_Q\right)$,  Eq.~\eqref{eq:rand_param} \Comment{a set of in. rand. param.}
        \State \quad -- ${\bm \rho}({\bf p})$, Eq.~\eqref{eq:dense}  \Comment{PDF for input variations}
        \State \quad -- $PCtype, PCorder$ \Comment{PC expansion parameters}
        \State \quad -- ${\bm w}, {\bf p}^{(k)}$, Eq.~\eqref{eq:nodes} \Comment{gener. weights \& points }
        
        \For{$k=1\ldots,K$}      
        \State {solve MVP$({\bf p}^{(k)})$ Eq.~\eqref{eq:strongf}}  \Comment{Quadrature points loop}
        \EndFor
        
        \For{${\bm i}=\{i_1,\ldots,i_Q\}$}      
        \State {$\widetilde{y}_{\bm i}\gets {\mathbb{E}[y({\bf p})\Phi_{\bm i}({\bf p})]}$, Eq.~\eqref{eq:disc_proj}}  \Comment{{discrete} projection}
        \EndFor
        \State Post-processing :
        \State \quad eval. $\mathbb{E}_{} \left[ {{ y}\left({{ {\bf p}}} \right)} \right], \mbox{Var}\left[ {{y}\left({{ {\bm p}}} 
    \right)} \right]$, Eq.~\eqref{eq:stat_moments} \Comment{statistics}
    
    \For{$q=1\ldots,Q$}      
        \State {eval. $\mathbb{E}\bigg{[}\frac{\partial y}{\partial p_q}\bigg{]},S_{q}$}, Eqs.~\eqref{mean_grad_sens}-\eqref{glob_sen}  \Comment{loc. \& glob. sens}
        \EndFor
    \end{algorithmic}
    \end{algorithm}

	\section{Shape Optimization}\label{sec:moso}

    In this section, the parametric MO shape optimization under uncertainties in terms of statistical moments is formulated. Therein, the global sensitivity is used to modify a scheme of the MO steepest descent algorithm. 

    \subsection{Objective functionals}
    Due to the geometrical uncertainties associated with ${\bf p}({\bm \xi})$ all the figures of merit \eqref{f1k}--\eqref{f6k} become functionals of the random geometry, denoted by $\Omega({\bf p}) \subset D$. In the context of classical optimal control theory~\cite{Pironneau1993}, especially its elaboration for optimization problems under uncertainties~\cite{Rosseel2012,Tiesler2012}, they can be represented in terms of certain target statistical quantities of interest, $F_l[\Omega(\bf p), {\bf H}]$ such as the expectation value, the linear sum of the mean and the variance value, the risk-aware probabilistic measure, the cumulative density function (CDF), see, e.g.,~\cite{Schillings2014}. 
    Correspondingly, the performance function of the QPR design can be represented as a functional $F_l(\Omega({\bf p}), {\bf H})$ of uncertain geometrical parameters ${\bf p}({\bm \xi})$, which is embedded in both the shape $\Omega$ and the state variable ${\bf H}:={\bf H}_{\Omega}$, i.e.  
    \begin{align}
    \label{eq:expect_obj}
        F_l(\Omega({\bf p}), {\bf H}):= \frac{1}{2}  \mathbb{E}\left[ \bigg{\|}f_l({\bf H}) -\overline{F}^0_l\bigg{\|}^2\,\right],
    \end{align}
    for $l=1,\ldots,L$, $L=3$ with $F_l(\Omega({\bf p}), {\bf H}) : H^1_0(D) \otimes L^2_{\rho}(\Gamma) \rightarrow \mathbb{R}$, which measure the distance between the objective functional $f_{l,3}({\cdot, {\bf H}})$ and the prescribed target value $\overline{F}_l^0$, in terms of the expectation value. 
    {This choice results from the observed measurement bias of the surface resistance for the third operating mode of the QPR, $n=3$, reported in~\cite{IEEEhowto:Keckert} as well as from the VBS analysis, which is shown in Figures~\ref{fig:f01}--\ref{fig:f03}. On the one hand, we aim at improving the accuracy/sensitivity of {the} measurement signal by maximizing the expectation of the focusing and homogeneity factors, that is, $F_1(\Omega({\bf p}), {\bf H})$ and $F_2(\Omega({\bf p}), {\bf H})$, respectively. On the other hand, the expectation of the dimensionless factor represented by $F_3(\Omega({\bf p}), {\bf H})$ needs to maximized, due to the unwanted heating of {the} normal-conducting flange.}

    Apparently, the finite assumption on the random field with $Q$ random variables, considered in Section~\ref{subsec:pf}, which is a result of the Doob–Dynkin Lemma~\cite{Babuska2004}, enables us to reformulate the stochastic shape optimization problem in a parametric MEP-constrained shape optimisation problem.
    Consequently, results from the deterministic optimization theory can be applied~\cite{Pironneau1993,Rosseel2012,Tiesler2012,putek2019} for the optimization of the QPR design under uncertainty.  
    
    \subsection{Shape optimization problem under uncertainties}
    
    The parametric MEP-constrained shape optimization problem of the QPR under uncertainties ${\bf p}({\bm \xi})$ aims to identify an optimal domain $\Omega^{*}({\bf p})$ as solution of  
    \begingroup\makeatletter\def\f@size{9.5}\check@mathfonts
    \begin{subequations} 
    \label{eq:smop}
    \begin{align}
        & \underset{{{{\Omega(\overline{\bf p})\, \in\, \mathbb{U}_{\rm add}}}}\,}{\mbox{inf }}   
          {{\bf F}}\left(\Omega({\bf p}), {\bf H}\right) = \left[{F}_1(\cdot),\;{F}_2(\cdot),\; {F}_3(\cdot)\, \right]^{\top}\label{eq:smopa}\\
        & \quad \mbox{s.t.}~ \mathbb{E}\left[ \int_{D}\,[{{\frac{1}{\mu_r}}\, (\nabla \times {\bf E})\cdot (\nabla \times {\bf v})} - \lambda\,  \epsilon_r\, {\bf E} \cdot {\bf v}]\,\mbox{d}{\xv}\right] = 0 
    \end{align}
    {for} ${{\bf H}:={{\bf H}_{\Omega} = \frac{1}{\omega \mu}} \nabla \times {\bf E}_{\Omega}}$ within a set of admissible shapes (parameters) 
    \begin{align}
        \label{eq:constr}
        \mathbb{U_{\rm add}} =\{{\bf p}({\bm \xi})\in \Pi\, \big{|}\, {\bf 0} \leq {\bf p}_{\rm min} \leq {\bf p} \leq {\bf p}_{\rm max} \},  
    \end{align}
    \end{subequations}
    \endgroup
    where for any $\Omega({\bf p}) \in \mathbb{U}_{\rm add}$ and ${\bf E}:={\bf E}_{\Omega}, {\bf v}:={\bf v}_{\Omega} \in V_{\rho}$ holds $\Omega({\bf p}) \subset D$. 
    
    Here, the inequality sign ${\leq}$ between vectors in \eqref{eq:constr} needs to be understood in a component-wise sense as follows: $ 0 \leq p_{{\rm min}_q} \leq \overline{p}_q \mp 3\cdot\sigma_q\leq p_{{\rm max}_q}, q=1,\ldots, Q$ with mean values $\overline{p}_q$ and standard deviations $\sigma_q$ of particular random input variables ${p_q}$. Furthermore, it is assumed that specific constraints $0 \leq p_{{\rm min}_q} \leq p_{{\rm max}_q}$ result from certain technological requirements. For instance, bounds of particular random geometrical parameters used in the VBS analysis are specified in {Table~\ref{l2ea4-t0}}. %
    
    In what follows, the focus is on using a gradient-based method~\cite{IEEEhowto:fliege2000} to solve the problem \eqref{eq:smop}. 
    To this purpose, a shape derivative needs to be provided.
    
    \subsection{Approximation of shape derivative}
    
    In the PDE-constrained shape optimization problem, the existence of a shape derivative has been proved by the Hadamard-Zalesio theorem, see, e.g.,  (\cite{bookDelfour2011}, Theorem 3.6). It states that a shape derivative for domains with smooth enough boundaries, can be represented as a distribution on the boundary, which depends only on the normal component of the perturbation
    \begin{align*}
       {\bm d} {\bf F}_l(\Omega; {\bf V}):=\int_{\partial D} h_l(\xv) {\bf V}(\xv)\cdot {\bf n}\, \mbox{d} {\xv}
    \end{align*}
    with $h(\xv)\in L^1(\partial D)$, where ${\bf V} : \mathbb{R}^d \rightarrow \mathbb{R}^d$ denotes the velocity field, while ${\bf n}$ is the outward normal unit vector. The distributed shape derivative $d F_l(\Omega; {\bf V})$ may also be expressed in a more general form as a volume integral over the whole domain~\cite{Laurain2016}. 
     
    Shape derivatives in stochastic settings are a topic of our ongoing research~\cite{tobepublished}. Apparently, under certain regularity conditions~\cite{Gatarek1992}, the shape derivative of functionals $F_l(\Omega({\bf p}), { \bf H}), \, l=1,\ldots,L$, defined by~\eqref{eq:expect_obj}, can be derived in the continuous framework using the velocity and adjoint variable methods as in~\cite{Komkov84, putek19}
    \begingroup\makeatletter\def\f@size{9.5}\check@mathfonts
    \begin{align}
        \label{eq:exactgradient}
        &{\bm d} {\bf F}_l(\Omega({\bf p}), {\bf V}) =\nonumber \\
        &\quad \mathbb{E}\bigg{[}\int_{\partial \Omega} \left(\frac{1}{\mu_{r1}} - \frac{1}{\mu_{r2}}\right)(\nabla \times {\bf E}_1) \cdot (\nabla \times {\bf E}_2)^{\top} ({\bf V}_q\cdot{\bf n})\, \mbox{d}{\bf x}\bigg{]} \nonumber \\ 
        &-\mathbb{E}\bigg{[}\int_{\partial \Omega} \lambda\,(\epsilon_{r1} - \epsilon_{r2}){\bf E}_1 \cdot {\bf E}_2^{\top} ({{\bf V}_q}\cdot{\bf n})\, \mbox{d}{\bf x}\bigg{]},  
    \end{align}
    \endgroup
    where the derivative of the boundary coordinate with respect to the $q$-th design variable is denoted by ${{\bf V}_q}:= \frac{\partial {\bf x}}{\partial p_q}$, $q=1\ldots,Q$. Here, ${\bf E}_1\in V_{\rho}$ and ${\bf E}_2\in V_{\rho}$ refer to the direct problem \eqref{eq:st_weak} and to a dual problem, respectively, which needs to be separately formulated for each objective functional and solved in order to calculate the shape derivative according to \eqref{eq:exactgradient}. 
    Moreover, when the effective Stroud-3 formula is used as a multi-dimensional quadrature rule, 2Q simulations of the deterministic problem \eqref{eq:st_weak} are needed to sufficiently approximate the probabilistic integrals \eqref{eq:disc_proj} and, in consequence, to find the statistical moments~\eqref{eq:stat_moments}. Thus, for the analyzed setup, $2QL = {30}$ simulations in every iteration of the MO optimization process are required to find the shape derivative, defined by {Eq.}~\eqref{eq:exactgradient}. But the Pareto front is approximated by $N={101}$ points. Hence, even when the steepest descent gradient method is used for solving \st{a} the MO optimization problem, {defined by Eq.~\eqref{eq:smop}}, this task {for the given setup} becomes time-consuming in the case of the 3D model of the QPR.

    For these reasons, in this work, the shape derivative~\eqref{eq:exactgradient} is approximated using {Eq.}~\eqref{mean_grad_sens} for $l=1,\ldots,L$,
    \begin{align}
    \label{eq:approxSD}
        {\bm d} {\bf F}_l(\Omega({\bf p})) \doteq  \sum\limits_{{\bm i}\in\mathcal{I}_{Q,P}} \tilde{g}_{l,\bm i} \frac{\partial \Phi_{\bm i}({\bf p})}{\partial { p}_q}\frac{\partial{\bf p}}{\partial \xi_q}\Bigg|_{p_q = {\overline{p}_q}}, 
    \end{align}
    with ${\bm d} {\bf F}_l(\Omega({\bf p}))\in \mathbb{R}^{Q}$ and $q=1\ldots,Q$. {This approximation results from} the Taylor's expansion-based approach for a deterministic measure of the robustness, developed, e.g., in~\cite{Harbrecht201091}.

    \subsection{Modified scheme of MO steepest descent}\label{moddified_schame}

    Since, a MO optimization problem of functionals competing with each other is considered, the concept of optimality needs to be replaced by the Pareto optimality framework.
    Accordingly, the solution of \eqref{eq:smopa} is said to be a set of optimal compromises in the Pareto sense, under following conditions~\cite{IEEEhowto:miettinen99}
    \begin{enumerate}
        \item $\Omega^{*}(\overline{\bf p}) =\Omega(\overline{\bf p}^{*}) \in \mathbb{U}_{\rm add}$ dominates $\Omega(\overline{\bf p}) \in \mathbb{U}_{\rm add}$, if ${\bm {{\bf F}}}(\Omega^{*}(\overline{\bf p}))\, {\leq}\, {\bm {{\bf F}}}(\Omega^{}(\overline{\bf p}))$ and ${\bm {{\bf F}}}(\Omega^{*}(\overline{\bf p})) \neq {\bm {{\bf F}}}(\Omega^{}(\overline{\bf p}))$, 
        \item $\Omega^{*}(\overline{\bf p}) \in \mathbb{U}_{\rm add}$ is called (globally) Pareto optimal, if there exists no $\Omega^{}(\overline{\bf p}) \in \mathbb{U}_{\rm add}$ dominating $\Omega^{*}(\overline{\bf p})$. 
    \end{enumerate}
    Here, a set of non-dominated point is called the Pareto set $\mathcal{P}_{\rm S}$, while its image is denoted as the Pareto front $\mathcal{P}_{F}$. 
    
    Now, to incorporate the global measure, i.e., the VBS analysis into a MO optimization flow, the {enhancement}
    gradient ${J} {\bf F}_l(\Omega({\bf p}))\in \mathbb{R}^Q$ for  $l=1\ldots,L$ is {introduced as} %
    \begin{align}
    \label{eq:enha_grad}
        {\bm J} {\bf F}(\Omega({\bf p})): = {\bf S}^{\top} \cdot [{\bm d} {\bf F}_1(\Omega({\bf p})),\ldots,{\bm d} {\bf F}_L(\Omega({\bf p}))].  
    \end{align}
    {This formulation} allows to benefit from both the local \eqref{eq:approxSD} and the global \eqref{glob_sen} sensitivity analysis. 
    {Thus,} in the modified scheme of the MO steepest descent algorithm, a matrix ${\bf S}$ with the VBS coefficients serves as a pre-conditioner. {It results in} speeding up the approximation of the Pareto front in terms of number of calling objective functions [cf. Table~\ref{Tab:mathexamp}]. 

    Furthermore, provided that some regularity conditions with respect to {Eq.}~\eqref{eq:smopa} are satisfied, it can be shown that the MO steepest descent method converges to a point satisfying the Karush-Kuhn-Tucker (KKT) conditions for Pareto optimality~\cite{IEEEhowto:fliege2000, desideri2012},
    such that
    \begin{align}
    \label{eq:kkt}
        \sum_{l=1}^L \alpha_l = 1,\,\, \textrm{and } \sum_{l=1}^L \alpha_l\, {J} {\bf F}_l(\Omega^{*}({\bf {p}})) = 0.
    \end{align}
    Therefore, the set of all the points, which satisfy {Eq.}~\eqref{eq:kkt} is called the set of substationary points $\mathcal{P}_{\rm S, sub}$. In contrary, if $\Omega^{}({\bf p}) \notin \mathcal{P}_{\rm S, sub}$, then there exits a descent direction ${\bf d} \in \mathbb{R}^Q$
    \begin{align}
    \label{eq:d}
        {\bf d}(\Omega^{}({\bf p})) = -\sum_{l=1}^L \alpha_l {J} {\bf F}_l(\Omega^{}({\bf {p}})),
    \end{align}
    such that  
    \begin{align}
    \label{eq:d}
        - {J} {\bf F}_l(\Omega^{}({\bf {p}}))^{\top} \cdot {\bf d}(\Omega^{}({\bf p}))\, {\geq}\, 0, \quad l=1,\ldots, L,
    \end{align}
    for which all the objectives are {non-increasing}. One way to determine a descent direction, which satisfies {Eq.}~\eqref{eq:d}, is to solve the auxiliary sub-optimization problem~\cite{schafner2002}. The other approach relies on analytically deriving ${\bf d}(\Omega^{}({\bf p}))$ by using the orthogonal projection~\cite{Liu2016,desideri2012}. For the convenience of the readers both methods are shortly described in Appendix~\ref{app:MO-descent-direction}.    

    \begin{algorithm}[H]
    \caption{MO descent direction under uncertainties} \label{alg:mosouu}
    \begin{algorithmic}[1]
        
        \Require : ${\Pi} = \left( {p}_1,\dots, {p}_Q\right)$, {Eq.}~\eqref{eq:rand_param}; ${\bm \rho}({\bf p})$, {Eq.}~\eqref{eq:dense}; $PCtype$, $PCorder$, ${\bm w}, {\bf p}^{(k)}$, {Eq.}~\eqref{eq:nodes}   \Comment{due to Algorithm~\ref{alg:psauq}}
        
        \State Set : initial point $\Omega({\bf p}^{<0>})$, 
        \State {\color{white}Set} : $precision$, ${\bm \kappa}^{<0>}$,
        \State {\color{white}Set} : $maxIters$, $maxCut$, $out$ 

        \For{$k=1,\ldots,maxIters$}
            \State set ncut = 0
            \State conduct UQ according to Algorithm~\ref{alg:psauq}
            \State evaluate gradients ${\bm J} {\bf F}(\Omega({\bf p}^{<k>}))$ using ~\eqref{eq:enha_grad}
 
            \State compute a direction ${\bf d}^{<k>}$ using either~\eqref{eq:smopaA} or~\eqref{eq:damalitic}
            
            \Repeat
                \State {$out\gets false$}
                \For{$l=1,\ldots,L$}
                    \If {$f_l(\overline{\bf p}^{<k>}+\kappa^{<k>}{\bf d}^{<k>}) \leq f_l(\overline{\bf p}^{<k>})$}
                        \State $\kappa^{<k>}\gets \kappa^{<k>}/2$
                        \State $ncut\gets ncut+1$
                        \State $out\gets true$
                        \State \textbf{break}
                    \Else
                        \State \textbf{continue} 
                    \EndIf
                    \If {$out$}
                    \State \textbf{break}
                    \EndIf
                \EndFor
                
                \State $\overline{\bf p}^{<k+1>}\gets\overline{\bf p}^{<k>}+\kappa^{<k>}{\bf d}^{<k>}$
                \For{$l=1\ldots,L$}
                    \State $f_l(\overline{\bf p}^{<k+1>}) \gets f_l(\overline{\bf p}^{<k>}+\kappa^{<k>}{\bf d}^{<k>})$
                \EndFor
                \State \State $\kappa^{<k+1>}\gets \min[{2\kappa^{<k>},\kappa^{<0>}}]$ \Comment{to restore step-size}
                \State \textbf{break}
                
            \Until{$ncut \leq maxCut$}
            
            \If {$\|\overline{\bf p}^{<k+1>}-\overline{\bf p}^{<k>} \|_2 \leq precision$}
                \State \textbf{stop}
            \EndIf
        \EndFor
    \end{algorithmic}
    \end{algorithm}

    Ultimately, when using a line search approach the MO steepest descent algorithm reads as   
    \begin{align}
        \label{eq:mosd}
        \Omega(\overline{\bf p}^{<k+1>}) = \Omega(\overline{\bf p}^{<k>} + \kappa^{<k>}{\bf d}^{<k+1>}), 
    \end{align}
    where the step length $\kappa^{<k>}>0$ is computed by an Armijo-like rule. Alternatively, the quadratic interpolation method can be used~\cite{IEEEhowto:fliege2000, Liu2016}.
    For a better clarity, the proposed method has been written as Algorithm~\ref{alg:mosouu} in the form of the pseudo-code. 
    
    {Finally, to speed up the Pareto front identification, in the 12-$th$ line of Algorithm~\ref{alg:mosouu}, the expectations appearing in the $if$ condition are replaced by their deterministic approximations, that is, $F_l(\cdot)\doteq f_l(\cdot),~l=1,\ldots, L$, which is motivated by the Taylor's expansion-based approach~\cite{Harbrecht201091}.}
    
    \section{Numerical Results \& Discussion}\label{sec:NumResults}
    
	The proposed algorithm has been verified using two test cases, which include an academic benchmark in the deterministic setting and the design of the QPR under uncertainties.
	
	\begin{table}[h]
    		\centering
    		\caption{Comparison of the efficiency for different bi-objective optimization methods. }
    		\label{Tab:mathexamp}
    \begin{tabular}{|l|c|c|c|c|}
    \hline
    Name & func.count & rel. error [\%]    & MSE(Y)    & MSE(X)    \\ \hline
    \textit{gamultiobj}()   &$31500$ & $-$ & $-$ & $-$\\ \hline
    \textit{paretosearch}() &$4351$ & 86.187 & 0.346 & 7.958 \\ \hline
    \textit{bsdm}()         &$4671$ & 85.171 & 0.335 & 8.179 \\ \hline
    \textit{bsdm-VBS}()     &$3368$ & 89.308 & 0.319 & 7.869 \\ \hline
    \end{tabular}
    \end{table}

    \subsection{Deterministic academic example}
    
    First, the efficiency of the proposed approach has been tested based on an academic benchmark problem{~\cite{report}}. It consists of the deterministic minimization problem of two functions $f_1({\bf x})$ and $f_2({\bf x})$ with ${\bf x}=(x,y)^{\top}$, which is defined as follows  
    \begin{align}
    \label{eq:det-examp}
        \max_{{\bf x}\in \mathbb{R}^2} {\bf {F}}({\bf x}) = \left[ -f_1({\bf x}), -f_2({\bf x}) \right]^{\mathsf{T}} 
    \end{align}
    with $f_1({\bf x})=4x^2+y^2+xy$ and $f_2({\bf x})=(x-1)^2+3(y-1)^2$, where ${\bf F}({\bf x}) : \mathbb{R}^2 \rightarrow \mathbb{R}^2$. For the solution of~\eqref{eq:det-examp} the Algorithm~\ref{alg:mosouu} is used with slight modifications, which allows us to compare its robustness with the already implemented methods to the MO optimization in MATLAB~\cite{matlab2018}. For this purpose a variant of the non-dominated sorting genetic algorithm II, the so-called \textit{gamultiobj} algorithm~\cite{deb:09} and the direct multisearch for multiobjective optimization method~\cite{custodio2011}, the so-called \textit{paretosearch} algorithm, have been used. In both solvers the standard settings have been applied. The Pareto front has been approximated using $N=300$ starting points. The efficiency of the methods has been measured with respect to the number of calling the objective functions, which has been summarized in Table~\ref{Tab:mathexamp}.     
    
    \begin{figure}[!tbh]
    	\centering
    	\includegraphics[width=\columnwidth]{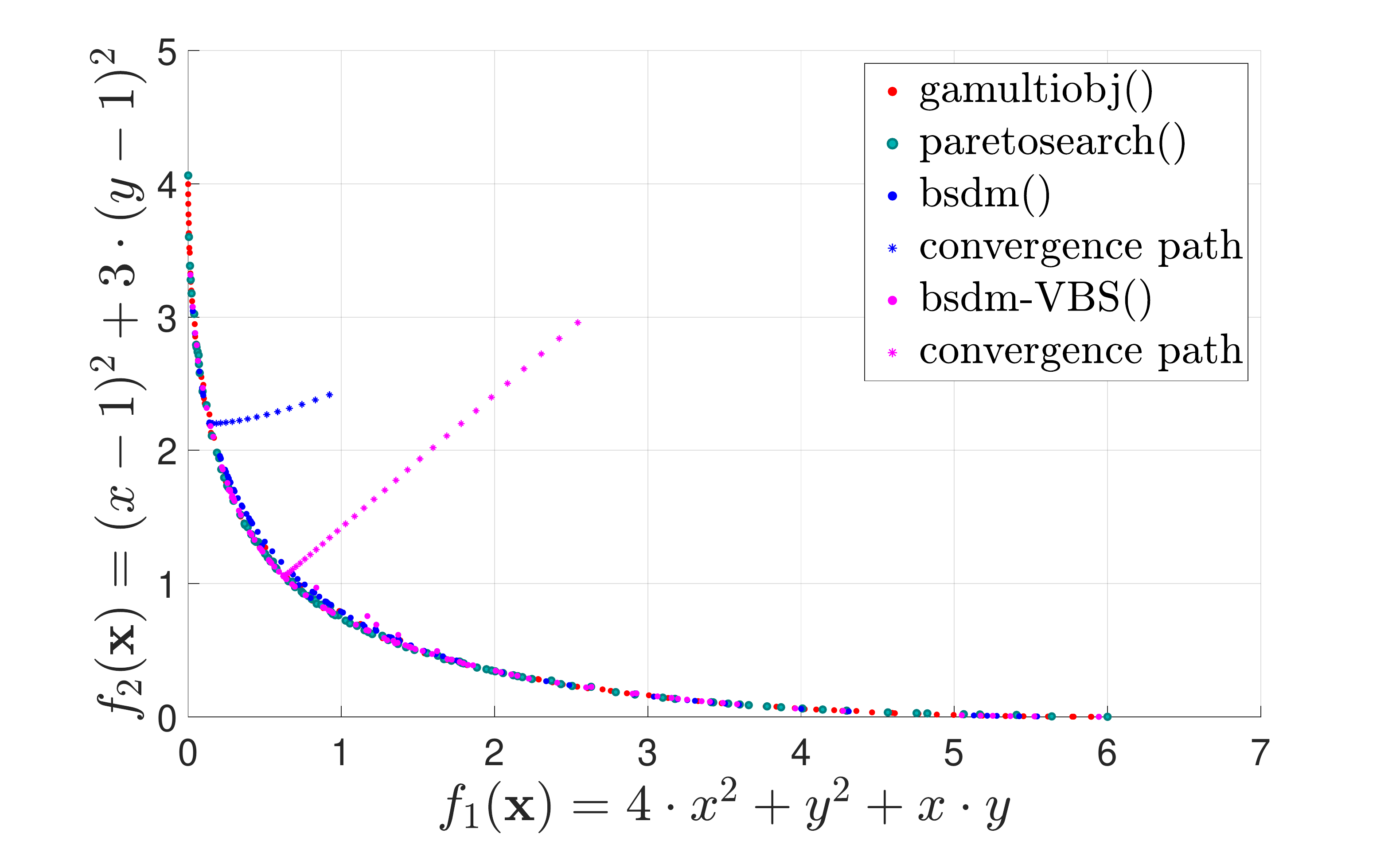}\hfill
    	\caption{Pareto front.} 
    	\label{fig:mathexPF}
    \end{figure}
        
    \begin{figure}[!tbh]
    	\centering
    	\includegraphics[width=\columnwidth]{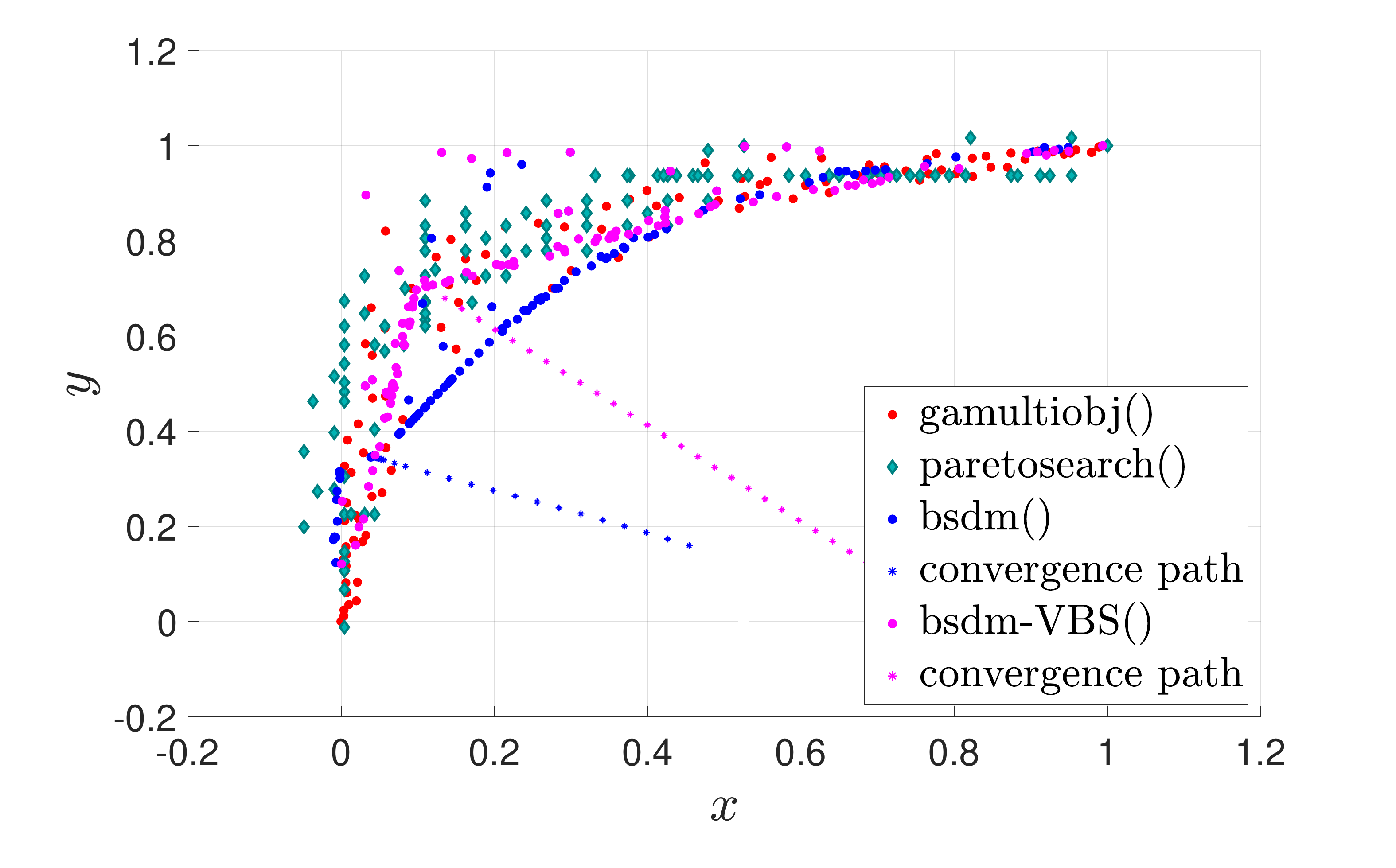}\hfill
    	\caption{Pareto set.} 
    	\label{fig:mathexPS}
    \end{figure}
    
    The deterministic counterpart of Algorithm~\ref{alg:mosouu} does not include the UQ algorithm. Therefore, based on the Taylor's expansion method, the VBS coefficients {Eq.}~\eqref{glob_sen} can be replaced by its deterministic approximation. Thus, assuming that the variables are independent~\cite{YAO2011450}, this decomposition is given by
    \begin{align}
    \label{eq:gs_approx}
    	     {\bf S} \doteq (\check{s})_{l,{q}}\in \mathbb{R}^{L,Q}, \quad \check{s}_{l,{q}} = \frac{\left(\frac{\partial f_l({\bf x})}{\partial x_q} \right)^2 \sigma_q^2}{\mbox{Var} [f_l({\bf x})] }, 
    \end{align}
    ${\text{for } q = 1,\ldots, Q},\; l =1\ldots,L.$, where the deterministic estimator of the total variance is defined as
    \begin{align*}
        \label{eq:mean-var}
        \mbox{Var} [f_l({\bf x})] := \sum_{q=1}^{Q} \left(\frac{\partial f_i({\bf x})}{\partial x_q} \right)^2 \sigma_q^2,\quad l=1,\ldots,L.  
    \end{align*}

    Here, $\sigma_q$ corresponds to the prescribed value of the standard deviation of each variable, usually as the ratio of its nominal value. Since the calculation of partial derivatives of $f_1({\bf x})$ and $f_2(\bf x)$ is straightforward, the enhanced gradient, when using {Eq.}~\eqref{eq:gs_approx}, can be easily found. Both the standard MO optimization methods, that is, the so-called \textit{bsdm} and \textit{bsdm-VBS} algorithms have been implemented in MATLAB. Furthermore, in order to initialize these MO optimizers, first the  
    {minima} of functions {$f_1$ and $f_2$}
    \begin{align*}
    &{\min_{{\bf x}\in \mathbb{R}^2} f_1({\bf x}_1)} = 9.6859\cdot10^{-10}, \\ &{\bf x}_1 = (-1.5096\cdot10^{-5},  -3.1274\cdot10^{-5})^{\top},\\
    &{\min_{{\bf x}\in \mathbb{R}^2} f_2({\bf x}_2)} = 9.6420\cdot10^{-10}, \\
    &{\bf x}_2 =(1.0000,  1.0000)^{\top}
    \end{align*}
    have been calculated. Next, initial points, $N=300$, have been generated using scaled and shifted \textit{rand(N,2)} values. Finally, the results for the Pareto front and set approximations have been depicted in Figures~\ref{fig:mathexPF} and~\ref{fig:mathexPS}, while the comparison in terms of the number of calling functions with the relative error in percentage and the mean squared error (MSE) are included in Table~\ref{Tab:mathexamp}. The MSE is defined as
    \begin{align*}
        MSE(y) = \frac{1}{N}\sum_{n=1}^N\left( y_n - \hat{y}_n\right)^2,
    \end{align*}
    where the reference solution is denoted by $\hat{y}_n$. In our study it is provided by the \textit{gamultiobj} algorithm. Likewise, in case of the relative error, the result for the \textit{gamultiobj} optimization is treated as the reference solution. It can be {concluded} that the  \textit{bsdm-VBS} method outperforms the more efficient MATLAB \textit{paretosearch} algorithm both in terms of the relative error by $22\%$ and the MSE$(\cdot)$ measure.

    {This mathematically oriented problem is considered here as a test case, which allows for demonstrating the applicability of the proposed approach in the more advanced QPR setting.}
    
    \begin{table}[!hbt]
       \centering
       \caption{Results for the 1D constrained optimization \\-- parameter domain.}
       \begin{tabular}{|l|c|c|c|}
           \hline
           \textbf{Name}  & $\Omega^*_1(\overline{{\bf p}})$ & $\Omega^*_2(\overline{{\bf p}})$ & $\Omega^*_3(\overline{{\bf p}})$ \\
           \hline
              $p_1$~{(gap)} $\qquad [\mathrm{mm}]$   & 0.50  & 1.49 &  0.51   \\ 
              $p_2$~{(rrods)} $~\quad [\mathrm{mm}]$ & 9.00  & 11.68  &  9.09  \\ 
              $p_3$~{(hloop)} $\,\quad [\mathrm{mm}]$ & 7.00  & 11.98 &  10.81   \\ 
              $p_4$~{(rloop)} $~\quad [\mathrm{mm}]$ & 6.00  & 5.99   &  4.00   \\
              $p_5$~{(wloop)} $\quad [\mathrm{mm}]$ & 43.92 & 43.98 &  36.08   \\
        \hline      
              $p_6$~{(dloop)} $\,\quad [\mathrm{mm}]$ & 4.00  & 4.00   &  4.05   \\
              $p_7$~{(rcoil)} $~~\quad [\mathrm{mm}]$ & 25.0  & 24.97   &  24.98   \\
              $p_8$~{(rsample)} $\,[\mathrm{mm}]$ & 35.0 & 35.00  &  35.03  \\
           \hline
       \end{tabular}
       \label{1Dopti}
    \end{table}

    \begin{table}[!hbt]
       \centering
       \caption{Results for the 1D constrained optimization \\-- objective space.}
       \begin{tabular}{|l|c|c|c|}
           \hline
           \textbf{Name}  & $\Omega^*_1(\overline{{\bf p}})$ & $\Omega^*_2(\overline{{\bf p}})$ & $\Omega^*_3(\overline{{\bf p}})$ \\
           \hline
              $F_1(\Omega^*(\overline{{\bf p}}), {\bf H})~[\mathrm{A^2/J}]$ & 9.56 $\cdot 10^7$  & 3.05 $\cdot 10^7$ & 6.44 $\cdot 10^7$ \\
              $F_2(\Omega^*(\overline{{\bf p}}), {\bf H})~[\mathrm{1/1}]$  & 0.174  & 0.23  & 0.12  \\
              $F_3(\Omega^*(\overline{{\bf p}}), {\bf H})~[1/1]$  & 2.47 $\cdot 10^6$   & 1.54 $\cdot 10^6$ & 5.20 $\cdot 10^6$  \\
           \hline
       \end{tabular}
       \label{1Dopti_fun}
    \end{table}

    \subsection{Optimization of the QPR}
    
    {Of utmost important objective functions in QPR optimization are the improvement of the resolution of {the} measurement signal by, on the one hand, maximizing eddy currents induced on the sample and, on the other hand, through the increase of the dissipated power at a given magnetic peak field. Additionally, the propagation of the magnetic field into the calorimetric chamber has to be minimized to limit the heating up of {the} normal-conducting flanges. This is in particular of crucial importance for the third mode where a measurement bias is observed in some experiments~\cite{Keckert2015}. Therefore, the optimization of the QPR design is formulated in terms of the focusing factor (\ref{f2k}), homogeneity factor (\ref{f3k}) and the dimensionless factor accounting for losses on the flanges (\ref{f6k}).}
    More precisely, the optimization problem is expressed by \eqref{eq:smop} with the expectation $F_1(\Omega({\bf p}), {\bf H})$, $F_2(\Omega({\bf p}), {\bf H})$ and $F_3(\Omega({\bf p}), {\bf H})$, defined by Eq.~\eqref{eq:expect_obj}, respectively. To solve this problem, the methodology described in Section~\ref{moddified_schame} is used. 
    
    In this respect, to initialize Algorithm~\ref{alg:mosouu}, first, the one dimensional (1D) optimization of $F_l(\Omega({\bf p}), {\bf H})$, $l=1,2,3$ with respect to parameters shown in Table~\ref{l2ea4-t0} needs to be carried out. Its results, which are included in Tables~\ref{1Dopti} and~\ref{1Dopti_fun} together with the VBS analysis, shown in Figs~\ref{fig:f01}, \ref{fig:f02} and~\ref{fig:f03}, allow for reducing
    quantities of interest 
    to $Q = 5$, ${\bf p}:=({p_1}, {p_2}, {p_3}, {p_4}, {p_5})^{\top}=(gap, rrods, hloop, rloop, wloop)^{\top}$. Hence, to approximate the Pareto front in the 3D objective space, the uniform random spread is applied. That is, the initial points, $N=101$, have been uniformly randomly generated using scaled and shifted $rand(N,Q)$ values between $\Omega^*_1(\overline{{\bf p}})$, $\Omega^*_2(\overline{{\bf p}})$ and $\Omega^*_3(\overline{{\bf p}})$ in $\mathbb{R}^Q$, listed in Table~\ref{1Dopti}. In addition, the initial step size, $\kappa^{<0>}$ = 0.5, the maximum number of step size and of  cuts at each iteration of the MO algorithm is set by ${maxCut} = 10$, ${maxIter} = 10$, while $precision =1\cdot10^{-5}$. Next, the backtracking method~\cite{NoceWrig06} has been used to approximate the length of the steepest gradient in every iteration, i.e, $\kappa^{<k+1>} = \kappa^{<k>}/2.0 $, for $k=1,\ldots,{maxCut}$.
    
   \begin{figure}[!tbh]
    	\centering
    	\includegraphics[width=\columnwidth]{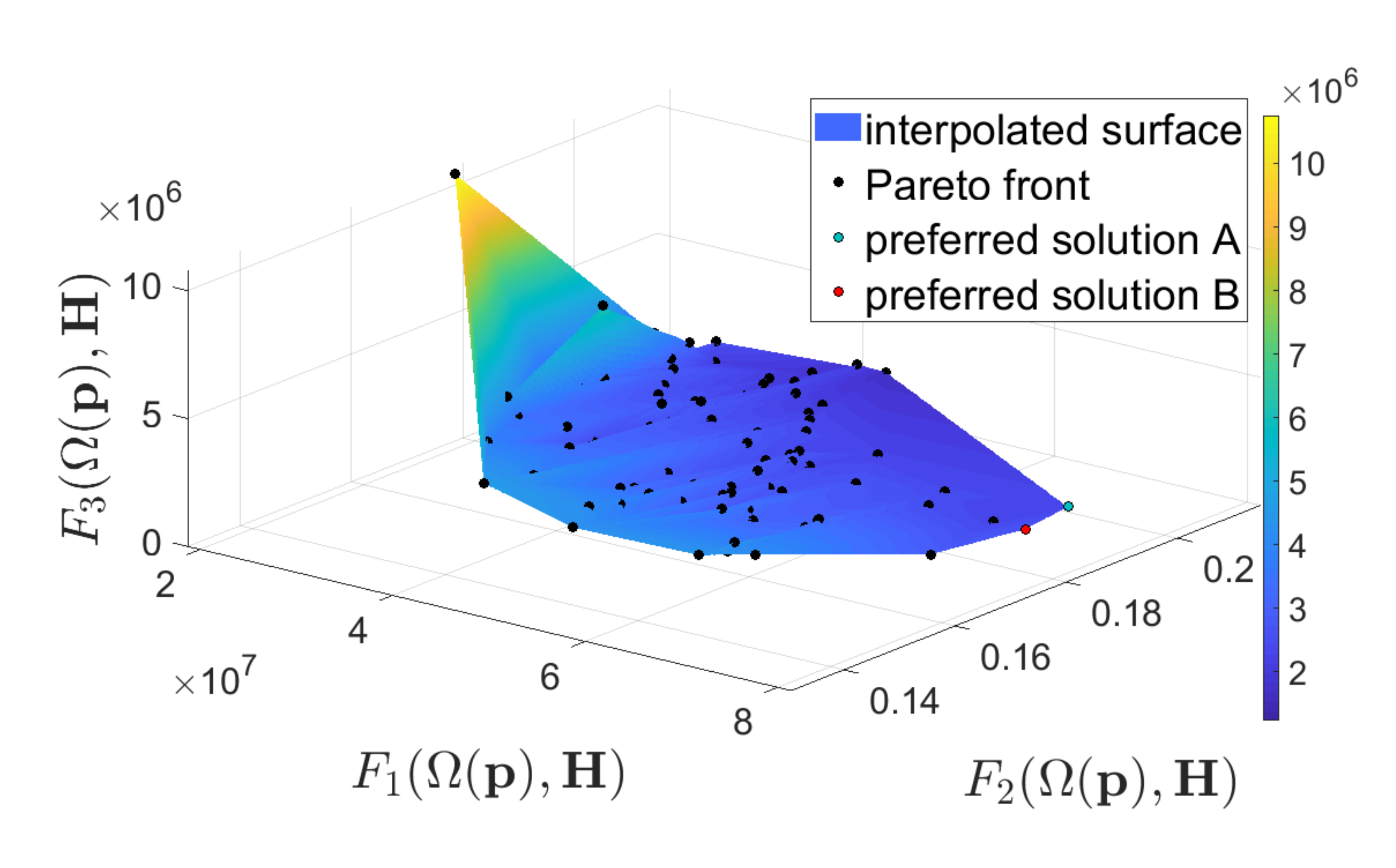}\hfill
    	\caption{Convergence to the Pareto front using the VBS-based approach~\eqref{eq:enha_grad} for several initial points $N=101$.
    	} 
    	\label{fig:PFQPRVBS}
    \end{figure}

    In the corresponding UQ model setup, these controllable geometrical parameters ${\bf p}$ are modeled with the PC. Here, the geometrical imperfections related to the manufacturing of the QPR of order $50-100~[\mathrm{\mu m}]$ are mimicked by the Gaussian distribution as follows
    \[
    p_q(\xi) = \overline{p}_q (1+\delta_q \cdot \xi_q), \quad q=1,\ldots, Q,
    \]
    where $\xi_q$ denotes the normally distributed variables and $\delta_q$ allows for controlling the magnitude of the perturbation regarding the production tolerance such that $\sigma_q := \delta_q \cdot \overline{p}_q = 0.05\,[\rm mm]$.
    
    Furthermore, according to Algorithm~\ref{alg:psauq} the appropriate polynomial type is chosen based on the Wiener-Askey scheme~\cite{Xiubook}. It corresponds to the input distribution, that is, the polynomials of the Hermite type. Moreover, the Stroud-3 formula is used to determine the Hermite polynomial coefficients, defined in Eq.~\eqref{eq:disc_proj}. Hence, exploiting the two-fold symmetry of the QPR, it requires $K:=2Q =10$ deterministic simulations of one quarter of the FE model expressed by Eq.~\eqref{eq:strongf} with $E_t=0$ on the symmetry plane using CST STUDIO SUITE\textsuperscript{\textregistered} for around {1 {million} tetrahedral mesh cells}. 
    
    The analysis discussed so far has to be performed for every initial {configuration} of the QPR. Hence, we have applied the MO steepest descent method with the {enhancement gradient}, defined by Eqs.~\eqref{eq:enha_grad}, for all the initially generated design points to be considered in the optimization problem. In particular, in Fig.~\ref{fig:PFQPRVBS}, the convergence of the MO steepest descent method to certain Pareto-stationary points {is} shown. 
    Additionally, the preferred configurations, the so called solutions $\rm A$ and $\rm B$, denoted by $\Omega^*_{\rm A}(\overline{{\bf p}})$ and $\Omega^*_{\rm B}(\overline{{\bf p}})$, have been listed in Tab.~\ref{MOconfABCD}. Moreover, for the comparison purpose, the configurations of the HZB-QPR $\Omega^*_{\rm HZB}(\overline{{\bf p}})$ {and the redesigned CERN-QPR $\Omega^*_{\rm CERN}(\overline{{\bf p}})$ are also included~\cite{IEEEhowto:Kleindienst,delPozoRomano:2678067}. The new version of the CERN-QPR is considered here in order to compare the optimized solutions with another existing QPR design.}
    
\begin{figure}[!tbh]
    	\centering
    	\includegraphics[width=1\columnwidth]{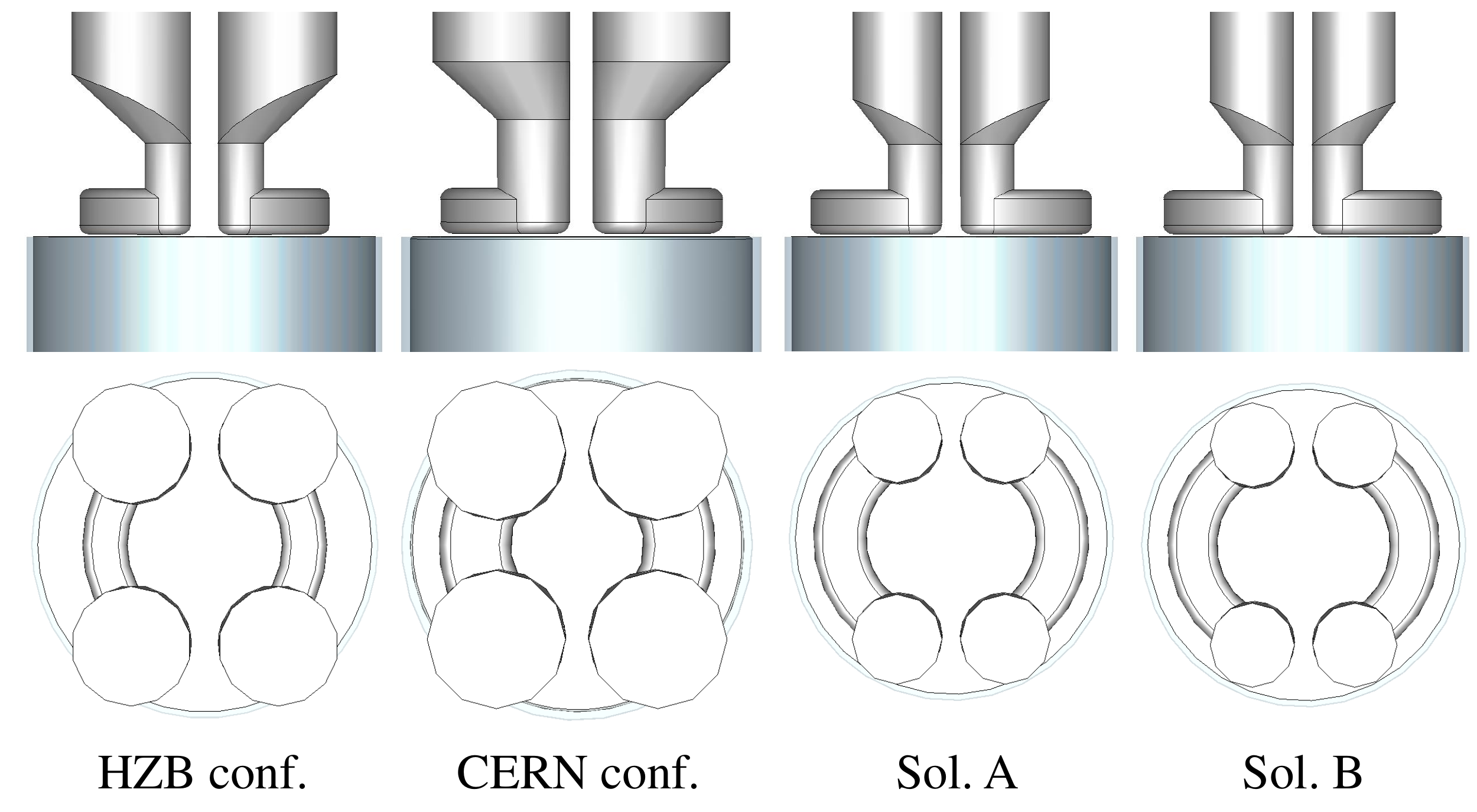}\hfill
    	\caption{Comparison between the shape of the HZB design, CERN design and the obtained solutions.} 
    	\label{fig:ShapeComparison}
    \end{figure}

\begin{figure}[!tbh]
    	\centering
    	\includegraphics[width=\columnwidth]{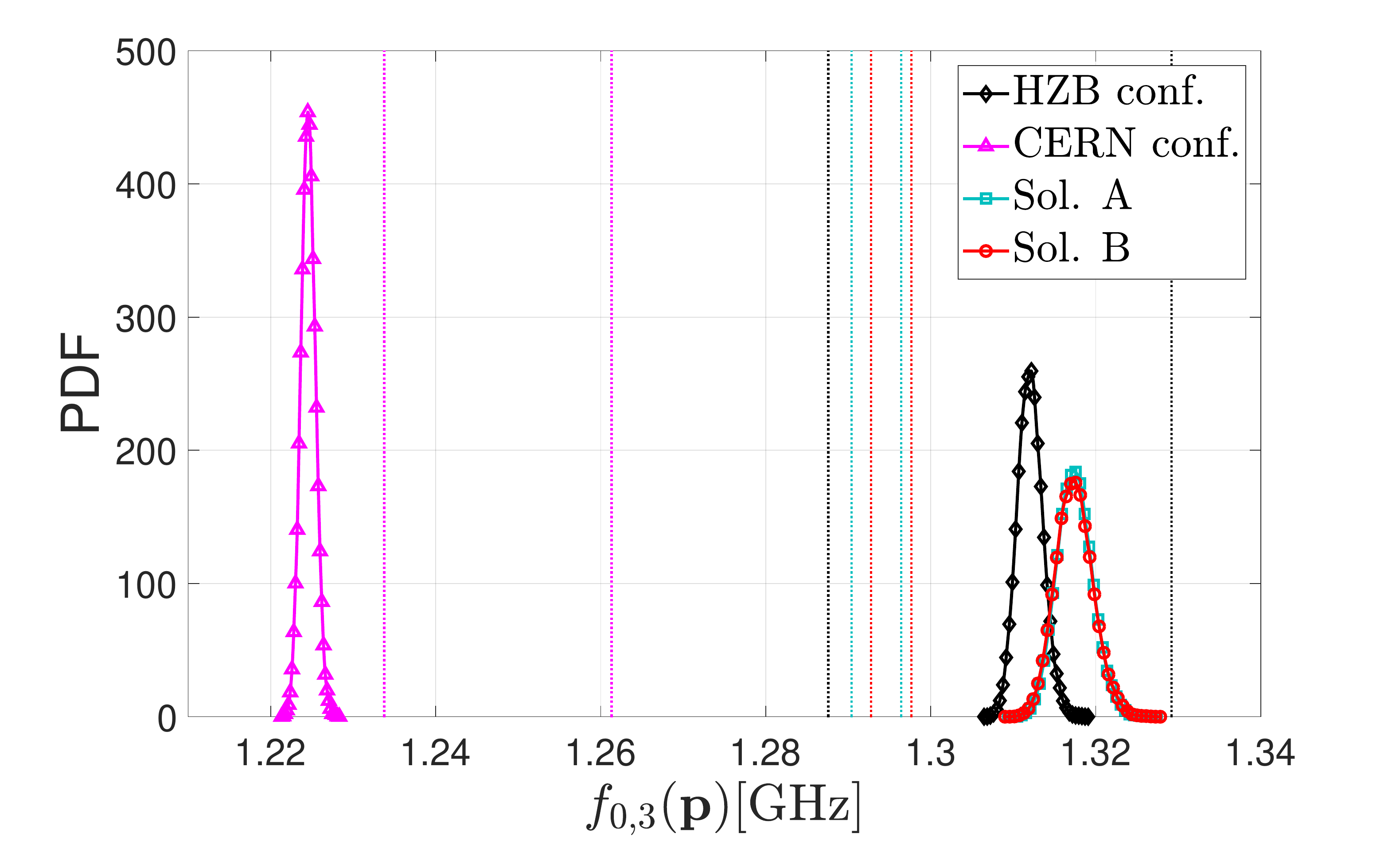}\hfill
    	\caption{{Probabilistic density function for the frequency of the third operating mode. The dashed vertical lines show the location of the next two closest neighboring modes calculated for the mean values given in Table~\ref{MOconfABCD}.
    	}    } 
    	\label{fig:pdff6}
    \end{figure}

   {The optimized geometries together with the HZB and CERN designs are depicted in Fig.~\ref{fig:ShapeComparison} and additionally summarized in Table~\ref{MOconfABCD}. The optimization results for the three operating modes in terms of objective functions incl. $F_0(\Omega^*(\overline{{\bf p}}), {\bf H})$, $F_4(\Omega^*(\overline{{\bf p}}), {\bf H})$ and $F_5(\Omega^*(\overline{{\bf p}}), {\bf H})$, are compared in Tabs.~\ref{MOopti_fm},~\ref{MOopti_sm} and~\ref{MOopti_tm}, respectively, for all analyzed designs. As can be seen, though, the optimization of the QPR has been conducted on the third mode, the objectives $F_1(\Omega^*(\overline{{\bf p}}), {\bf H})$, $F_2(\Omega^*(\overline{{\bf p}}), {\bf H})$ and $F_3(\Omega^*(\overline{{\bf p}}), {\bf H})$ have been significantly improved for all (first/second/third) operating modes by {$11-15/28-33/49-55\,[\%]$, $40-49/37-44/35-42\,[\%],$ and $136-165/156-189/217-260\,[\%]$}, respectively. This improvement in the first and second operating modes should be treated as a side effect of optimizing the geometry with respect to the third mode.  
   
   The inspection of these tables revealed also that some serious issues with the new CERN-QPR design might be caused by the new pole shoes, for which the field within the coaxial gap becomes relatively large. This, in turn, can lead to the increase of the unwanted heating of {the} normal-conducting flange [cf. $F_3(\Omega^*(\overline{{\bf p}}), \cdot)$ for the QPR designs, listed in Tabs.~\ref{MOopti_fm},~\ref{MOopti_sm} and~\ref{MOopti_tm}]. It can be noticed that for the proposed design (solutions A and B) the dimensionless factor $F_3(\Omega^*(\overline{{\bf p}}), \cdot)$ takes averagely 2.5 times larger values than in the case of the {HZB} design and more than 5 times bigger compared to the CERN-QPR configuration. Another explanation for the larger measurement bias of the surface resistance is that the CERN-QPR modes and the neighboring dipole modes are close to each other for  a perfectly aligned QPR design. Thus, even the relatively small input deviations of geometrical variables related to the 
   {manufacturing} tolerances can yield the excitation of the neighboring modes. The difference in frequency of the closest neighboring mode to the third operating mode is \unit[9.2]{MHz} for the CERN-QPR and \unit[17.2]{MHz} for the HZB-QPR while it is \unit[20.9]{MHz} and \unit[19.9]{MHz} for Sol. A and Sol. B, respectively [cf. Fig.~\ref{fig:pdff6}]. In our case, the found QPR configurations (A and B) have larger separation between the third mode and its neighboring mode. }

\begin{figure}[!tbh]
    	\centering
    	\includegraphics[width=\columnwidth]{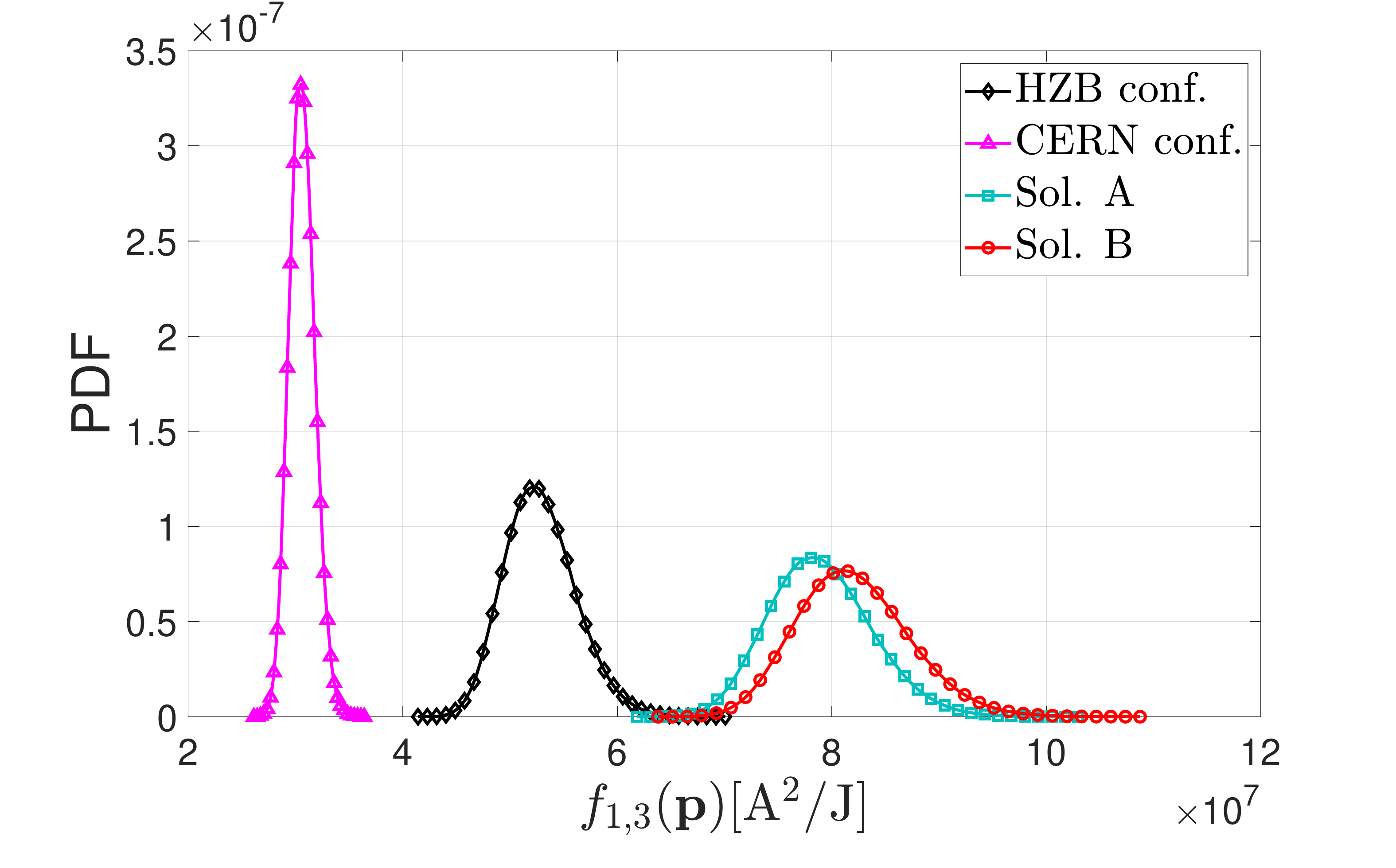}\hfill
    	\caption{Probabilistic density function for the focusing factor.} 
    	\label{fig:pdff1}
    \end{figure}

\begin{figure}[!tbh]
    	\centering
    	\includegraphics[width=\columnwidth]{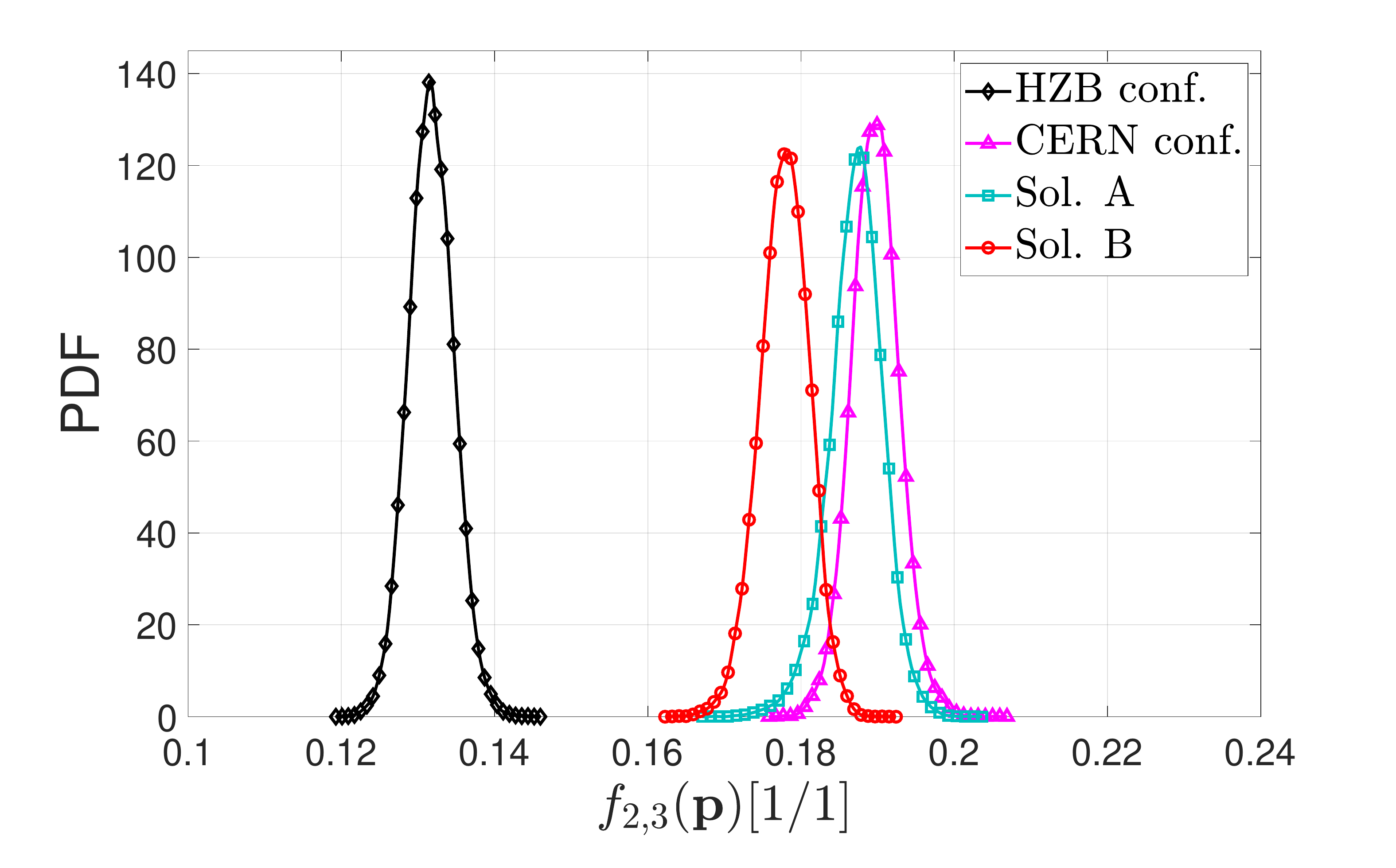}\hfill
    	\caption{Probabilistic density function for the homogeneity factor.} 
    	\label{fig:pdff2}
    \end{figure}

\begin{figure}[!tbh]
    	\centering
    	\includegraphics[width=\columnwidth]{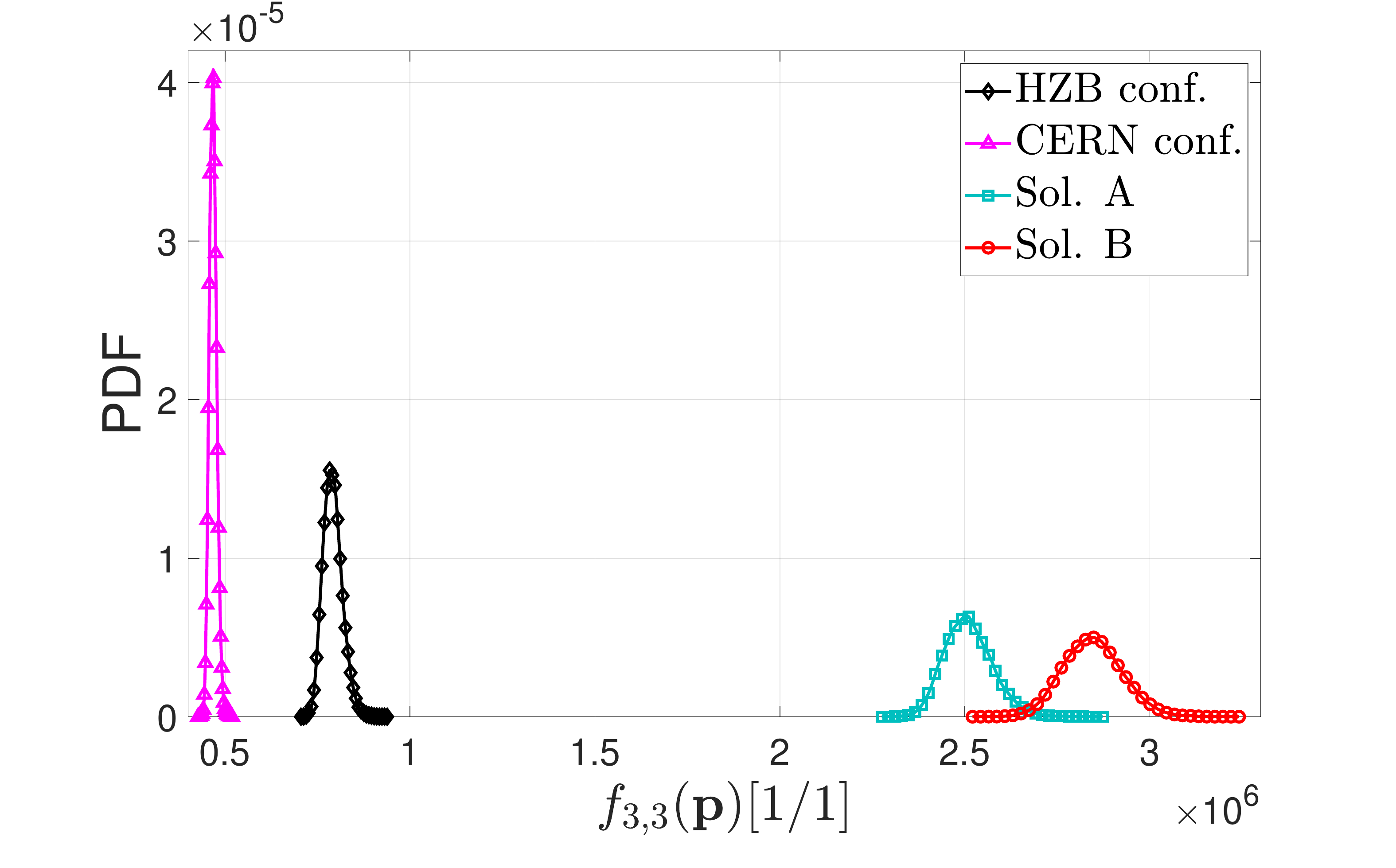}\hfill
    	\caption{Probabilistic density function for the dimensionless factor {representing the losses on {the} normal-conducting flange}.} 
    	\label{fig:pdff3}
    \end{figure}

\begin{figure}[!tbh]
    	\centering
    	\includegraphics[width=\columnwidth]{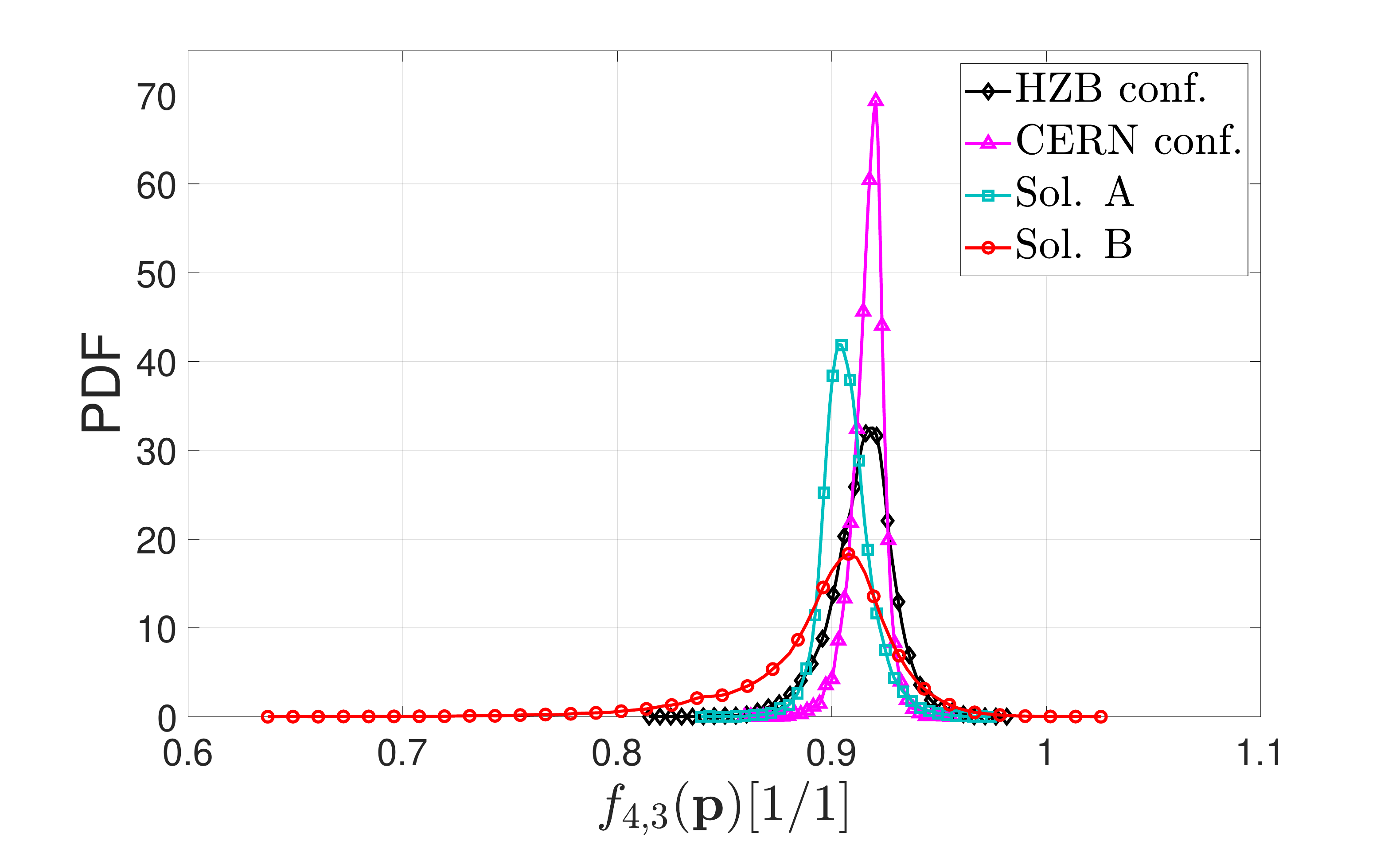}\hfill
    	\caption{Probabilistic density function for the ratio of the magnetic {peak} values.} 
    	\label{fig:pdff4}
    \end{figure}

\begin{figure}[!tbh]
    	\centering
    	\includegraphics[width=\columnwidth]{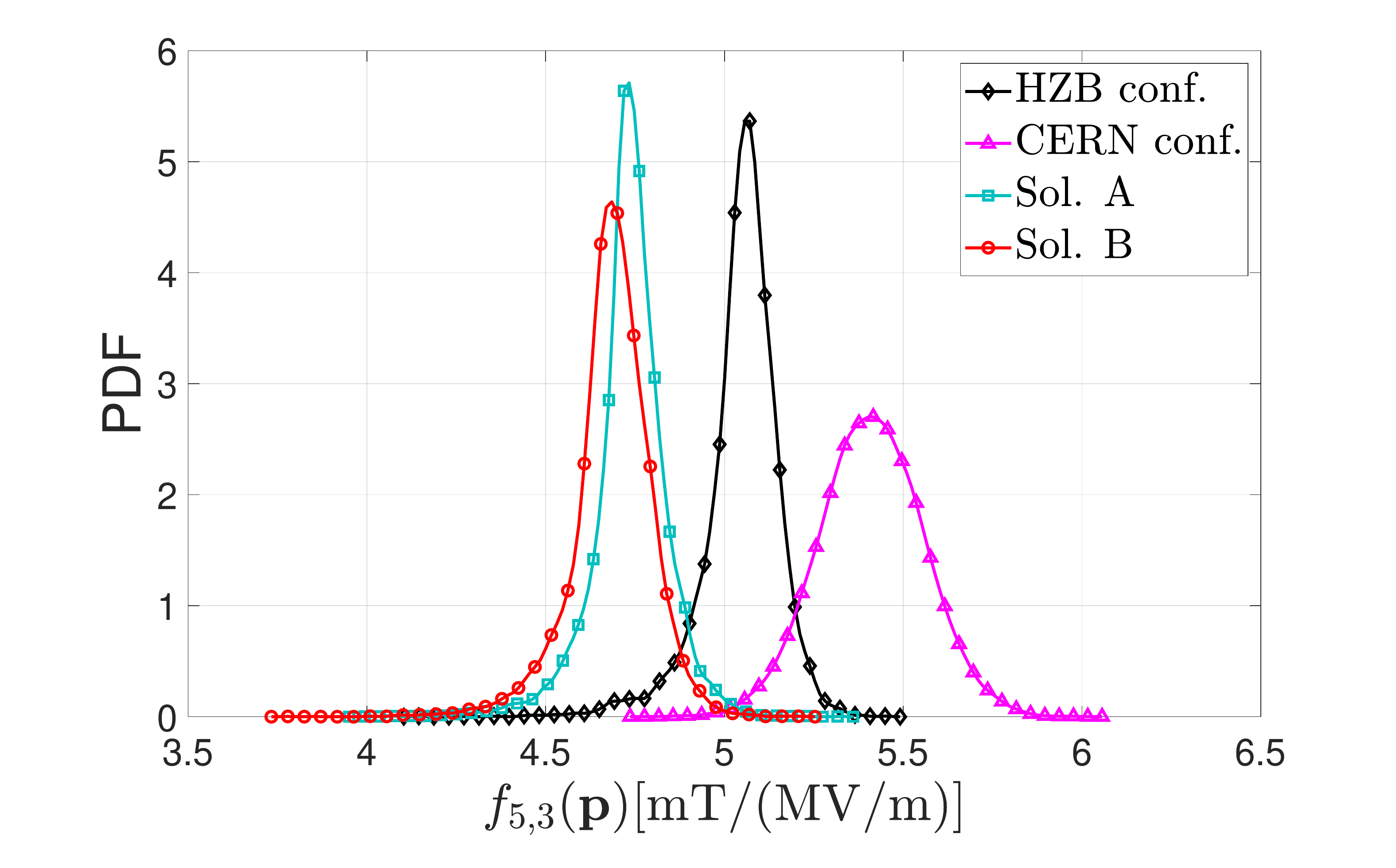}\hfill
    	\caption{Probabilistic density function for the {ratio} of the magnetic {peak} by the electric {peak} values.} 
    	\label{fig:pdff5}
    \end{figure}

   One can also notice that $F_4(\Omega^*(\overline{{\bf p}}), {\bf H})$ and $F_5(\Omega^*(\overline{{\bf p}}), {\bf H})$ have been reduced for the solution A and B by {$0.6-1.0/1.2-1.7/0.8-1.9\,[\%]$, $37-39/26-28/6-7\,[\%]$} compared to the 
   {HZB} structure. {In the case of the CERN design, the corresponding functionals have been changed by $-0.4/-0.9/0.3\,[\%]$, $-33/-29/7\,[\%]$ with respect to the {HZB} design. However, it should be noticed that} these objectives, that is, $F_4(\Omega^*(\overline{{\bf p}}), {\bf H})$ and $F_5(\Omega^*(\overline{{\bf p}}), {\bf H})$ have not been a subject of the MO optimization. In fact, the QPR is supposed to be operated at low field values (in comparison with accelerating cavities) and, therefore, the field emission problem, i.e, $F_5(\Omega^*(\overline{{\bf p}}), {\bf H})$ is not a very critical issue~\cite[p.~70]{IEEEhowto:Kleindienst}. In the end, the small deviation of the operating modes about {$2.2/1.4/0.4\,[\%]$}, which is observed, can be compensated by changing the length of the rods and/or $lrans1$ and $ltrans2$ parameters~\cite{IEEEhowto:Kleindienst}, depicted in Fig.~\ref{fig:QPRCrossSection}. Finally, the PDF's analysis of the shapes associated with the final configurations is depicted in {Figs.~\ref{fig:pdff1} -- \ref{fig:pdff6}}, respectively. They are obtained  by  evaluating $1\cdot10^4$ times the corresponding truncated response surface models, defined by Eq.~\ref{eq:trun}. 

    \renewcommand{\arraystretch}{1.5}
    \begin{table*}[!hbt]
       \centering
       \caption{Results for the MO optimization -- parameter domain~\footnote{{For practical reasons, the 
       {resulting} optimized parameters are given with the required accuracy (the second decimal point).} {In addition to the pole-shoe, the resonator body of CERN-QPR and HZB-QPR have different shapes. The most important differences are the radius and the length of the cylindrical resonator body which are, respectively, \unit[105]{mm} and \unit[354]{mm} for the CERN-QPR and \unit[120]{mm} and \unit[304.75]{mm} for the HZB-QPR. Additionally, the coaxial gap considered for the CERN-QPR and HZB-QPR are \unit[2]{mm} and \unit[1.5]{mm}, respectively. }}}
       \begin{tabular}{|C{2.8cm}|C{1.5cm}|C{1.5cm}||C{1.5cm}|C{1.5cm}|}
           \hline
           \textbf{Name}  & $\Omega^*_{\rm HZB}(\overline{{\bf p}})$ & $\Omega^*_{\rm CERN}(\overline{{\bf p}})$ & $\Omega^*_{\rm A}(\overline{{\bf p}})$ & $\Omega^*_{\rm B}(\overline{{\bf p}})$ \\
           \hline
              $p_1$~{(gap)} $\qquad [\mathrm{mm}]$ & 0.50 & 0.70 & 0.58  & 0.55      \\ 
              $p_2$~{(rrods)} $~\quad [\mathrm{mm}]$ & 13.00 & 15.00 & 9.76  & 9.14  \\ 
              $p_3$~{(hloop)} $\,\quad [\mathrm{mm}]$ & 10.00 & 10.00 & 9.72  & 9.64 \\ 
              $p_4$~{(rloop)} $~\quad [\mathrm{mm}]$ & 5.00 & 8.00 & 5.92  & 5.56    \\
              $p_5$~{(wloop)} $\quad [\mathrm{mm}]$ & 44.00 & 40.93 & 43.79 & 43.53  \\
        \hline      
              $p_6$~{(dloop)} $\,\quad [\mathrm{mm}]$ & 6.00 & 5.00 & 4.00  & 4.00   \\
              $p_7$~{(rcoil)} $~~\quad [\mathrm{mm}]$ & 22.408 & 23.00 & 25.00  & 25.00    \\
              $p_8$~{(rsample)} $\,[\mathrm{mm}]$ & 37.50 & 37.50 & 35.0 & 35.00        \\
           \hline
       \end{tabular}
       \label{MOconfABCD}
    \end{table*}
    \renewcommand{\arraystretch}{1.25}
    
\renewcommand{\arraystretch}{1.5}
    \begin{table*}[!hbt]
       \centering
       \caption{Results of the MO optimization for the first mode -- objective space~\footnote{The columns  with percentage $[\%]$ indicate a ratio (increase +/decrease -) of optimized configurations to $\Omega^*_{\rm HZB}(\overline{{\bf p}})$.}}
       \begin{tabular}{|C{4.2cm}|C{1.3cm}|C{1.6cm}|C{1.0cm}||C{1.1cm}|C{1.0cm}|C{1.1cm}|C{1.0cm}|}
           \hline
           \textbf{Means/Configurations} & $\Omega^*_{\rm HZB}(\overline{{\bf p}})$ & $\Omega^*_{\rm CERN}(\overline{{\bf p}})$  &[\%]& $\Omega^*_{\rm A}(\overline{{\bf p}})$ & [\%]& $\Omega^*_{\rm B}(\overline{{\bf p}})$ & [\%]  \\
           \hline
              $F_1(\Omega^*(\overline{{\bf p}}), \cdot)~[\mathrm{M~A^2/J}]$ & 50.07   & 32.15 & -36.55 & 56.31  &  11.13 & 58.47 & 15.39  \\
              $F_2(\Omega^*(\overline{{\bf p}}), \cdot)~[\mathrm{1/1}]$     & 0.155   & 0.218 & 41.15  & 0.227  &  48.84 & 0.216 & 39.70  \\ 
              $F_3(\Omega^*(\overline{{\bf p}}), \cdot)~[\mathrm{M~1/1}]$   & 1.668   & 0.890 & -46.64 & 3.941  &  136.3 & 4.421 & 165.1  \\
              \hline 
              $F_4(\Omega^*(\overline{{\bf p}}), \cdot)~[1/1]$              & 0.910   & 0.906 & -0.43  & 0.901  &  -1.01  & 0.905 & -0.62  \\
              $F_5(\Omega^*(\overline{{\bf p}}), \cdot)~[\mathrm{mT/(MV/m)}]$ & 7.888 & 5.250 & -32.93 & 4.824  &  -38.84 & 4.940 & -37.38  \\
              $F_0(\Omega^*(\overline{{\bf p}}), \cdot)~[\mathrm{GHz}]$     & 0.429   & 0.398 & -7.21  & 0.439  &  2.21   & 0.439 & 2.23  \\ 
           \hline
       \end{tabular}
       \label{MOopti_fm}
    \end{table*}
    \renewcommand{\arraystretch}{1.25}

\renewcommand{\arraystretch}{1.5}
    \begin{table*}[!hbt]
       \centering
       \caption{Results of the MO optimization for the second mode -- objective space~\footnote{The columns with percentage $[\%]$ indicate a ratio (increase +/decrease -) of optimized configurations to $\Omega^*_{\rm HZB}(\overline{{\bf p}})$.}}
       \begin{tabular}{|C{4.2cm}|C{1.3cm}|C{1.6cm}|C{1.0cm}||C{1.1cm}|C{1.0cm}|C{1.1cm}|C{1.0cm}|}
           \hline
           \textbf{Means/Configurations} & $\Omega^*_{\rm HZB}(\overline{{\bf p}})$ & $\Omega^*_{\rm CERN}(\overline{{\bf p}})$  &[\%]& $\Omega^*_{\rm A}(\overline{{\bf p}})$ & [\%]& $\Omega^*_{\rm B}(\overline{{\bf p}})$ & [\%]  \\
           \hline
              $F_1(\Omega^*(\overline{{\bf p}}), \cdot)~[\mathrm{M~A^2/J}]$  & 48.36  & 29.36 & -39.28  & 61.86  &  27.93 & 64.30 & 32.97  \\
              $F_2(\Omega^*(\overline{{\bf p}}), \cdot)~[\mathrm{1/1}]$      & 0.146  & 0.207 & 42.05  & 0.211  &  44.44  & 0.200 & 36.93  \\ 
              $F_3(\Omega^*(\overline{{\bf p}}), \cdot)~[\mathrm{M~1/1}]$    & 1.293  & 0.705 & -45.47  & 3.322  &  156.8 & 3.742 & 189.3  \\
              \hline 
              $F_4(\Omega^*(\overline{{\bf p}}), \cdot)~[1/1]$               & 0.920  & 0.911 & -0.94 & 0.904  &  -1.71   & 0.909 & -1.24  \\
              $F_5(\Omega^*(\overline{{\bf p}}), \cdot)~[\mathrm{mT/(MV/m)}]$ & 7.289 & 5.13  & -29.61  & 5.251  & -27.95 & 5.368 & -26.35 \\
              $F_0(\Omega^*(\overline{{\bf p}}), \cdot)~[\mathrm{GHz}]$      & 0.867  & 0.807 & -6.95 & 0.879  &  1.39    & 0.879 & 1.4  \\ 
           \hline
       \end{tabular}
       \label{MOopti_sm}
    \end{table*}
    \renewcommand{\arraystretch}{1.25}

\renewcommand{\arraystretch}{1.5}
    \begin{table*}[!hbt]
       \centering
       \caption{Results of the MO optimization for the third mode -- objective space~\footnote{The columns  with percentage $[\%]$ indicate a ratio (increase +/decrease -) of optimized configurations to $\Omega^*_{\rm HZB}(\overline{{\bf p}})$.}}
       \begin{tabular}{|C{4.2cm}|C{1.3cm}|C{1.6cm}|C{1.0cm}||C{1.1cm}|C{1.0cm}|C{1.1cm}|C{1.0cm}|}
           \hline
           \textbf{Means/Configurations} & $\Omega^*_{\rm HZB}(\overline{{\bf p}})$ & $\Omega^*_{\rm CERN}(\overline{{\bf p}})$  &[\%]& $\Omega^*_{\rm A}(\overline{{\bf p}})$ & [\%]& $\Omega^*_{\rm B}(\overline{{\bf p}})$ & [\%] \\
           \hline
              $F_1(\Omega^*(\overline{{\bf p}}), \cdot)~[\mathrm{M~A^2/J}]$   & 52.28 & 30.63 & -42.05 & 78.98  &  49.43  & 82.04 & 55.21 \\
              $F_2(\Omega^*(\overline{{\bf p}}), \cdot)~[\mathrm{1/1}]$       & 0.132 & 0.19  & 44.00  & 0.187  &  42.09  & 0.178 & 35.0\\ 
              $F_3(\Omega^*(\overline{{\bf p}}), \cdot)~[\mathrm{M~1/1}]$     & 0.791 & 0.467 & -40.89 & 2.501  &  217.4  & 2.846 & 259.9\\
              \hline 
              $F_4(\Omega^*(\overline{{\bf p}}), \cdot)~[1/1]$                & 0.914 & 0.917 & 0.3    & 0.907  &  -0.81  & 0.897  & -1.94\\
              $F_5(\Omega^*(\overline{{\bf p}}), \cdot)~[\mathrm{mT/(MV/m)}]$ & 5.048 & 5.411 & 7.19   & 4.736  &  -6.18  & 4.685 & -7.19\\
              $F_0(\Omega^*(\overline{{\bf p}}), \cdot)~[\mathrm{GHz}]$       & 1.312 & 1.225 & -6.67  & 1.317  &   0.41  & 1.317 & 0.41\\ 
           \hline
       \end{tabular}
       \label{MOopti_tm}
    \end{table*}
    \renewcommand{\arraystretch}{1.25}

    \section{Conclusion} \label{sec:cons}
    
    In our work, we applied the PC and the VBS analysis to find {a} robust design of the QPR. The expected value of {the} figures of merit has been chosen as a robust measure.  Following the VBS decomposition, on the one hand, the set of quantities of interest has been reduced. On the other hand, the coefficients of the VBS have been used to construct the { enhancement}
    gradient. This way, the scheme of the MO steepest descent method has been modified in order to take into account the { manufacturing} tolerance related to the geometrical parameters. 
 
    Furthermore, based on the technical specification of the QPR, the preferred {solutions have}
    been chosen {as} $\rm A, B$ to find the optimized QPR configuration. As can be seen in Tables~\ref{MOopti_fm},~\ref{MOopti_sm} and~\ref{MOopti_tm}, the robust designs of the QPR (solutions $\rm A, B$) allow for increasing the focusing factor of the third mode by 50-57~\% and 158-168~\%  in comparison with the HZB and CERN designs, respectively. The focusing factor of the first and second modes are also improved in parallel as a side effect by 12-17~\% and 29-35~\%, respectively. This gives rise to a better resolution in the {determination of the} surface resistance in different frequencies. Additionally, the dimensionless factor of the third mode, which takes into account the propagation of the magnetic field into the coaxial gap around the sample, is more than twice bigger than for the HZB and CERN configuration. This improvement helps to decrease the measurement bias observed for the third mode in HZB and CERN designs. These results can be further improved using both the Pareto technique and the robust/reliability based frameworks, where either the expectations and standard deviations or the probability of failure of objective functionals are considered. It is seen as a promising direction for future investigations.

	\begin{acknowledgments}
	
    This work has been supported by the German Federal Ministry for Research and Education BMBF under contract 05H18HRRB1. {Furthermore, the authors would like to thank Dr. S.~Keckert and Dr.~O. Kugeler (Helmholtz-Zentrum Berlin, Germany) for the QPR-related discussions.} 
    
	\end{acknowledgments}
	
\appendix{\color{black}

\section{Methods for MO descent direction}
\label{app:MO-descent-direction}

As in~\cite{IEEEhowto:fliege2000, desideri2012} the problem of computing a steepest descent direction can be formulated as 
a convex quadratic problem with linear inequality constraints
\begin{align}
\label{eq:smopaA}
    \underset{{\bm{\alpha}}\in \mathbb{A}_{\rm add}}{\mbox{min }}   
      \Big{\|} \sum_{l=1}^L \alpha_l {J} {\bf F}_l(\Omega^{}({\bf {p}}))\Big{\|}^2,
\end{align}
where $\mathbb{A}_{\rm add} \subset \mathbb{R}^{L}$ is a set of the admissible vectors $\bm \alpha$ with 
{non-negative} components and ${\bm \alpha}^{\top}\cdot {\bm \alpha}=1$ such that
\begin{align*}
    \mathbb{A}_{\rm add} = \Big{\{} {\bm \alpha} \in \mathbb{R}^{L} \Big{|}\, \alpha_l \geq 0, \forall l = 1,\ldots, L, ,\,\, \sum_{l=1}^L \alpha_l = 1  \Big{\}}
\end{align*}
Then, ${\bf d}$ is defined by \eqref{eq:d}. 

The solution of \eqref{eq:smopa} can be also derived analytically using the Theorem 2 in~\cite{Liu2016}. Please note that for 
simplicity, the dependence of $(\Omega({\bf p}))$ has been suppressed in the following equations and $\nabla f_l:={ {\bm J}}{\bf F}_l(\Omega^{}({\bf {p}}))$. Let us define~\cite{Liu2016,liu2016online} 
\begingroup\makeatletter\def\f@size{9.5}\check@mathfonts
    \begin{align}
        {\bf  D}_{t,{i_1},{i_2},\ldots,t} = \sum_{j=1}^{t-1}\,y_j(-\nabla f_{i_j})-\nabla f_{i_t}, \nonumber
    \end{align}
    such that
    \begin{align}
    \label{eq:damalitic}
        {\bf d}_{t,{i_1},{i_2},\ldots,t} = \frac{{\bf  D}_{t,{i_1},{i_2},\ldots,t}}{\| {\bf  D}_{t,{i_1},{i_2},\ldots,t}\|}, 
    \end{align}
    which satisfies
    \begin{align}
        \nabla f_{i_1}^{\top} {\bf  d}_{t,{i_1},{i_2},\ldots,t} = \nabla f_{i_2}^{\top} {\bf  d}_{t,{i_1},{i_2},\ldots,t} = ...= \nabla f_{i_t}^{\top} {\bf  d}_{t,{i_1},{i_2},\ldots,t}. \nonumber  
    \end{align}
    \endgroup
    Then, if $\{ \nabla f_1,\ldots, \nabla f_n \}$ is {a} linearly independent set and ${\bf  d}$ is the analytic solution to \eqref{eq:smopa}, there {exists} a positive integer $p$ with $1 \leq r \leq n$ such that ${\bf  d} = {\bf  d}_{r,{i_1},{i_2},\ldots,r}$ is defined by \eqref{eq:damalitic}, where $\{i_1,\ldots, i_r\} \subset \{1,\ldots, n\}$. In addition, ${\bf  d} = {\bf  d}_{r,{i_1},{i_2},\ldots,r}$ is also the $r$ objective steepest descent direction for $f_{i_1},\ldots, f_{i_r}$.
}
	
	\bibliography{main}
	
\end{document}